\documentclass[aps,
prd]{revtex4}

\usepackage{latexsym}
\usepackage{epsfig}
\usepackage{scrextend}

\def\be{\begin{equation}}
\def\ee{\end{equation}}
\def\bea{\begin{eqnarray}}
\def\eea{\end{eqnarray}}
\def\ba{\begin{array}} 
\def\ea{\end{array}}
\def\bc{\begin{center}}
\def\ec{\end{center}}

\def\ghost#1{}

\def\simge{\mathrel{%
   \rlap{\raise 0.511ex \hbox{$>$}}{\lower 0.511ex \hbox{$\sim$}}}}
\def\simle{\mathrel{
   \rlap{\raise 0.511ex \hbox{$<$}}{\lower 0.511ex \hbox{$\sim$}}}}

\def\dis{\displaystyle}

\def\be{\begin{equation}}
\def\ee{\end{equation}}
\def\bea{\begin{eqnarray}}
\def\eea{\end{eqnarray}}

\def\ba{\begin{array}} 
\def\ea{\end{array}}
\def\bc{\begin{center}}
\def\ec{\end{center}}

\def\ghost#1{}

\def\simge{\mathrel{%
   \rlap{\raise 0.511ex \hbox{$>$}}{\lower 0.511ex \hbox{$\sim$}}}}
\def\simle{\mathrel{
   \rlap{\raise 0.511ex \hbox{$<$}}{\lower 0.511ex \hbox{$\sim$}}}}

\begin{document}

\title{THE \,SUPERSYMMETRIC \,STANDARD \,MODEL  \vspace{8mm}\\

\it with a Brout-Englert-Higgs boson as spin-0 partner of the $Z$ \vspace{1cm}\\}

\vspace{-3mm}

\author{Pierre FAYET\vspace{5mm}
\\}

\vspace{5mm}

\address{Laboratoire de Physique Th\'eorique de l'ENS
{\small \ (UMR 8549 CNRS)} \\
 24 rue Lhomond, 75231 Paris Cedex 05, France \vspace{3mm}\\}

\begin{abstract}

\textwidth 18cm
Supersymmetric extensions of the standard model  lead us to expect superpartners for all particles, 
\hbox{spin-0} squarks and sleptons and \hbox{spin-$\frac{1}{2}$}
gluinos, charginos and neutralinos, with an odd $R$-parity making the lightest one stable.

\vspace{.8mm}
The electroweak breaking is induced  by a pair of spin-0 doublets, leading  to several charged and neutral \hbox{BE-Higgs} bosons. 
These theories also lead to gauge/Higgs unification  by providing spin-0 bosons as {\it extra states for spin-1 gauge bosons} 
within massive gauge multiplets.

\vspace{1mm}
 
In particular, the 125 GeV$/c^2$ boson recently observed at CERN, most likely a BE-Higgs boson associated 
with the electroweak breaking, may also be interpreted, up to a mixing angle induced by supersymmetry breaking, as
{\it the spin-0 partner of the $Z$} under {\it two} supersymmetry transformations.

\vspace{1mm}
We also discuss how the compactification of extra dimensions, relying on $R$-parity and other discrete symmetries,
may determine both the grand-unification and supersymmetry-breaking scales.

\end{abstract}

\maketitle

\vspace{.2cm}

\section{Introduction}

Is there a ``superworld'' of new particles\,?
Could half of the particles at least have escaped our observations\,?
Do new states of matter of exist\,? 
After the prediction of antimatter by Dirac, supersymmetric extensions of the standard model
lead to anticipate the possible existence, next to quarks and leptons,
of associated \hbox{spin-0} {\it squarks} and {\it sleptons},
with the gluons, $W^\pm,\,Z$ and photon also associated with new superpartners,
{\it gluinos, charginos\,} and {\it neutralinos} \cite{R,ssm,ff,fayet79}.
These new states are characterized by a quantum number called {\it $R$-parity} related to baryon and lepton numbers, 
obtained from a  discrete remnant of a continuous $U(1)_R$ symmetry acting chirally on the supersymmetry generator,
broken to $R$-parity by the gravitino and gluino masses  \cite{grav,glu}.
\vspace{1.5mm}

The spontaneous breaking of the electroweak symmetry is induced, in contrast with the Standard Model \cite{ws},
by {\it a pair of spin-0 doublets} responsible for charged-lepton and down-quark masses, and 
up-quark masses, respectively \cite{R,ssm}. 
 This leads  to expect {\it  charged \hbox{spin-0} bosons $H^\pm$}, and additional neutral ones. 
Such theories possess many attractive features, providing in particular a natural place for fundamental spin-0 bosons next to spin-1 
and spin-$\frac{1}{2}$ particles, and the possibility of associating spin-1 with spin-0 particles within massive gauge multiplets of supersymmetry.
And we keep waiting 
for  signs of superpartners  \cite{susyat,susycms}  and additional Brout-Englert-Higgs bosons \cite{hat,hcms},
beyond the one recently found at the CERN LHC \cite{higgs,higgs2}.

\vspace{1.5mm}

This new boson with a mass close to 125 GeV$/c^2$ may actually be interpreted (up to a mixing angle, possibly small,  induced by supersymmetry breaking) as
{\it a spin-0
partner of the spin-1 $\,Z$} within a massive gauge multiplet of supersymmetry,
providing within a theory of electroweak and strong interactions the first example of {\it two known fundamental particles 
of different spins related by supersymmetry} -- in spite of their different electroweak properties.

\vspace{1.5mm}

We shall review here the main steps followed in the construction of the Supersymmetric Standard Model,
parallely to related developments in $N=2$ and $N=4$ supersymmetric theories. These more speculative 
theories may also be expressed using extra compact dimensions, 
that may play an essential role in the breaking of the  supersymmetry and grand-unification symmetries at the compactification scale(s) \cite{56}.
We also refer the reader to the standard review articles \cite{rev,rev2,rev3,rev4}, 
and leave more detailed discussions on the present status of supersymmetric theories, 
including the role of neutralinos 
as possible dark matter candidates
and the effect of the new particles on gauge-coupling unification, to subsequent contributions to this volume.

 \section{Relate bosons with fermions, yes, but how\,?}

To begin with,  according to common knowledge, supersymmetry relates, or should relate, bosons with fermions:
\be
\underbrace{\hbox{Bosons}}_{\hbox{\footnotesize integer spin}} \ \ \stackrel{\stackrel{\hbox{\footnotesize SUSY}}{}}{\longleftrightarrow} \ \ 
\underbrace{\hbox{Fermions}}_{\hbox{\footnotesize half-integer spin}}\!\!\!.
\ee
But can such an idea be of any help in understanding the real world 
of particles and interactions\,? Could one relate, for example, mesons with baryons,
\be
\label{mb}
\hbox{Mesons}\ \ \longleftrightarrow \ \ \hbox{Baryons} \ \ ?
\ee
as attempted by Miyazawa in the sixties \cite{mia} within a non-relativistic framework\,? 
Or in a more modern way dealing now with fundamental particles, can one relate the bosons, messengers of interactions,
with the fermions, constituents of matter, to arrive at some sort of  unification
\be
\label{unif}
\hbox{Forces}\ \ \longleftrightarrow \ \ \hbox{Matter} \ \ ?
\ee
This would be very attractive, but unfortunately
things don't work out that way.

\vspace{2mm}

Indeed it turns out that supersymmetry should associate known bosons with new fermions, and known fermions with new bosons.
While this it now often presented as obvious, it was long taken, and even mocked, as a sign of the irrelevance of supersymmetry.
Still, part of the utopic association (\ref{unif}) between forces and matter may turn out to be relevant in the case of {\it dark matter}, for which supersymmetric theories provide 
a natural candidate in connection with $R$-parity conservation (cf.~(\ref{fdm},\ref{fdm2}) in Sec.\,\ref{sec:ssm}).
\vspace{2mm}

The supersymmetry algebra 

\vspace{-4mm}
\be
\label{alg}
\left\{ \  
\begin{array}{ccc}
\{ \ Q\, , \, {\bar Q} \ \} \!&=&\! 
- \, 2\,\gamma_{\mu}   P^{\mu} \, \vspace {2mm} \cr 
[ \ Q\,, \, P^{\mu} \,] \!&=& \ 0\ \ 
\end{array}  \right.      
\ee
relates supersymmetry transformations with space-time translations. It  was introduced in the years 1971-73 \cite{gl,va,wz,ra} with various motivations,  including:
is it at origin of parity non-conservation \cite{gl},
or is the neutrino a Goldstone particle \cite{va}\,?
More interestingly, the intimate connection of supersymmetry with space-time translations implies
that a theory invariant under local supersymmetry transformations must include general relativity, leading to supergravity theories
\cite{vs,sugra,sugra2}.

\vspace{2mm}

However,  even knowing about the mathematical existence of such an algebra, 
with bosonic and fermionic degrees of freedom jointly described using
superfields \cite{sf},
fundamental bosons and fermions do not seem to have much in common.
And it  is hard to imagine how they could be related 
by a spin-$\frac{1}{2}$ generator, in a relativistic theory.
Beyond the obvious fact that bosons and fermions have different masses, to which we shall return later, 
the gauge bosons, mediators of interactions, and the quarks and leptons  do not have the same gauge quantum numbers.

\vspace{2mm}

In addition supersymmetric gauge theories \cite{wz2,ym}  systematically involve spin-$\frac{1}{2}$ Majorana fermions, unknown in Nature (with a possible exception for neutrinos in case lepton number turns out not to be exactly conserved). In contrast known fundamental fermions, quarks and leptons, correspond to Dirac spinors carrying conserved quantum numbers, baryon number $B$ and lepton number $L$. These are even known, or were known in the past, as 
{\it fermionic numbers}, to emphasize that they are carried by fundamental fermions only, not by bosons.
Of course this no longer appears as necessary today, now that we got familiar with supersymmetric extensions of the standard model 
and ready to accept the possible existence of spin-0 bosons, squarks and sleptons, carrying $B$  and $L$ almost by definition \cite{ssm}, 
but this was  once viewed as quite a heretic hypothesis. Furthermore, just attributing $B$ and $L$ to squarks and sleptons 
does not necessarily guarantee
that these quantum numbers are going to be conserved, at least to a sufficiently good approximation. This is also where $R$-symmetry and $R$-parity 
are going to play an essential role \cite{ssm,ff,fayet79}.

\vspace{2mm}

Altogether  supersymmetry first seemed irrelevant to the description of the real world, and many physicists kept this point of view for quite some time.

\section{General \,features}
\label{sec:general}

\vspace{-.3mm}

\subsection{The specificities of spontaneous supersymmetry breaking}
\label{sub:spec}

\vspace{-.3mm}

There is also the difficult question of how to obtain a spontaneous breaking of the supersymmetry. This is by far not trivial owing to the specificities of its algebra, allowing to express the hamiltonian from the squares of the four components of the supersymmetry generator, as
\be
\label{h}
H=\ \frac{1}{4}\ \sum_\alpha\ Q_\alpha^2\,.
\ee
It implies that {\it a supersymmetric vacuum state must have a vanishing energy}, which first seemed to prevent any spontaneous breaking of supersymmetry to possibly occur \cite {iz}.  In any case such a breaking  should lead to a massless spin-$\frac{1}{2}$ Goldstone fermion, unobserved. 

\vspace{2mm}

Nevertheless, in spite of this apparently general argument, spontaneous supersymmetry breaking turned out to be possible, 
although it is very severely constrained. Indeed in global supersymmetry, instead of simply trying to make a supersymmetric vacuum state 
unstable,
as one would normally do for any ordinary symmetry, 
one has to arrange for such a symmetric state to be {\it totally absent}, 
as it would otherwise have vanishing energy and be stable owing to the relation (\ref{h}) between the hamiltonian and supersymmetry generator.

\vspace{2mm}

Such a very special situation may be obtained 
using either a mechanism relying on a $U(1)$ gauge group and associated $\xi D$ term included in the Lagrangian density~\cite{fi}.
Or using an appropriate set of chiral superfields including at least a gauge singlet one 
with its corresponding $\,\sigma F$ term from a linear term in the superpotential
\cite{F1,or}, interacting through a suitable superpotential carefully  chosen {\it with the help of an $R$ symmetry} \cite {R}.
 These models also lead to the systematic existence  of {\it classically-flat directions} of the potential
 (valleys)  associated 
 with classically-massless particles 
 ({\it moduli} or {\it pseudomoduli\,}), in connection with the fact that one makes it impossible for all auxiliary components 
 to vanish simultaneously; and this in a generic way, thanks to the use of an $R$ symmetry.
 
\vspace{2mm}

In most situations on the other hand, spin-0 fields generally tend, in order to minimize the energy, 
to adjust their vacuum expectation values so that all auxiliary fields vanish simultaneously, with supersymmetry remaining conserved. 
At the same time the other symmetries may well be, quite easily,  spontaneously broken, including charge and color gauge symmetries 
if some charged-slepton or squark fields were to acquire non-vanishing v.e.v.'s. This would be, also, a real disaster\,! 
In practice one will always have to pay sufficient attention to the supersymmetry-breaking mechanism, 
so that {\it all} 
\,squarks and sleptons (and charged BE-Higgs bosons) acquire large  positive mass$^2$ (i.e.\,{\it ``no tachyons''}), 
and the vacuum state preserves electric charge and color as required, 
avoiding charge-or-color-breaking (CCB) minima.

\subsection{Gluino masses, and metastable vacua}
\label{subsec:meta}

Note that {\it metastable vacuum states} with a very long lifetime may have to be considered, 
separated by a potential barrier from a lower-energy stable minimum of the energy, for which charge or color symmetries 
could be spontaneously broken. This may the case, in particular, in the presence of additional spin-0 gluon fields 
introduced in  \cite{glu} to turn gluinos into Dirac particles, with an underlying motivation from extended supersymmetry
\cite{hyper,24}.  

\vspace{2mm}
Gluinos would remain massless in the presence of an unbroken continuous $R$ symmetry, also denoted $U(1)_R$, acting chirally on them. 
Gluino mass terms, however,  may be generated radiatively from their Yukawa couplings  to a new set of massive 
{\it messenger-quark superfields vectorially coupled to gauge superfields}, sensitive both to the source of super\-symme\-try-breaking
(e.g.~through auxiliary-component v.e.v.'s $\,<\!F\!>$ or $<\!D\!>\,$), and to a source of $R$-symmetry breaking, for Majorana gluinos  \cite{glu}. It is however difficult to generate radiatively large gluino masses, unless one accepts to consider really very large masses for messenger quarks, as frequently done now.

\vspace{2mm}
One can also generate in this way a Dirac gluino mass term, which preserves 
the continuous $R$-symmetry.
The new spin-0  gluon fields introduced to turn gluinos into Dirac particles, now called ``sgluons'',  tend however to acquire radiatively-generated  negative mass$^2$ 
from their couplings to messenger quarks 
\cite{glu} (cf.~footnote {\small 5}), so that the corresponding desired vacuum state must be stabilized in order 
to avoid color breaking.

\vspace{2mm}

This may be done by adding in the Lagrangian density a direct gauge-invariant chiral-octet-superfield mass term, 
breaking explicitly the $U(1)_R$ symmetry down to $R$-parity. It includes, next to a ``sgluon'' mass$^2$ term, a direct
$\Delta R= \pm 2\,$  gluino Majorana mass term for the second octet of ``paragluinos'' breaking the continuous $U(1)_R$. 
This Majorana mass term  splits the Dirac gluino octet
into two Majorana mass eigenstates
through the {\it see-saw mechanism for Dirac gluinos}. This one is formally analogous to the see-saw mechanism for neutrinos
that became popular later. While the color-preserving vacuum, with massive gluinos, gets locally stabilized in this way, it is
only {\it metastable}~\cite{glu} 
(cf.~footnote {\small 6}), 
an interesting feature compatible with phenomenological requirements  also occurring in other situations,
which attracted some attention later~\cite{iss}.

\vspace{2mm}

Let us return to Majorana gluinos. Their mass terms are not forbidden
in the supergravity framework where the gravitino acquires a mass $m_{3/2}$ so that $R$-symmetry gets reduced to $R$-parity,
allowing for direct gaugino mass terms \cite{grav}, that may  be generated from gravity-induced supersymmetry breaking.
Jointly with the direct higgsino mass term $\mu$ (or effective mass term $\mu_{\rm eff}$), these terms allow for {\it both charginos\,} to be {\it heavier than $m_W$}, 
as is now necessary \cite{cfg,pdg}.

\subsection{The fate of the Goldstone fermion, and  related interactions of a light gravitino}

A massless Goldstone fermion appears in spontaneously broken globally-supersymmetric theories, 
which is in principle viewed as an embarrassment. 
This Goldstone fermion, however, may be eliminated by the super-Higgs mechanism within supergravity theories 
\cite{vs,sugra,sugra2,crem}. Still it
may actually survive 
\vspace{-.3mm}
under the form of 
the $\pm\,\frac{1}{2}$ polarisation states of a massive but possibly very light spin-$\frac{3}{2}$ gravitino.
But a very light gravitino still behaves as a (quasi-massless) 
\hbox{spin-$\frac{1}{2}$} goldstino according to the equivalence theorm of supersymmetry \cite{grav},
in which case we get back to our starting point, still having to discuss the fate of the Goldstone fermion\,!

\vspace{2mm}
Thanks to $R$-parity, however, this Goldstone fermion, being $R$-odd,  has no direct couplings to ordinary particles only. 
It couples bosons to fermions with the multiplets of supersymmetry, i.e.~ordinary particles to superpartners  (as yet unseen),
in a way fixed by the boson-fermion mass spectrum through the supercurrent conservation equation.
Furthermore, its interactions may be much weaker than weak interactions, if the supersymmetry-breaking scale parameter
($\sqrt d$ or $\sqrt F$), related to the gravitino mass by $m_{3/2}=\kappa d/ \sqrt 6=\kappa F/ \sqrt 3$ with $\kappa^2=8\pi G_N$,   \,is sufficiently large \cite{fayet79,grav}. Supersymmetry is then said to be broken ``at a high scale'',
the \hbox{spin-$\frac{1}{2}$} polarisation states of such a  light gravitino behaving as an ``almost-invisible'' 
goldstino.

\vspace{2mm}
The gravitino is then the lightest supersymmetric particle, or LSP, 
with a very-weakly-interacting \cite{weak} and thus early-decoupling gravitino appearing as a possible candidate for the non-baryonic dark matter 
of the Universe \cite{gravdm}.
With such a light gravitino LSP, the next-to-lightest supersymmetric particle (NLSP), usually a neutralino, is expected to decay,
possibly with a long lifetime, according to 
\be
\label{decaygrav}
\hbox{\it neutralino} \ \rightarrow\  \gamma\ +\ \hbox{\it unobserved gravitino},
\ee 
leading to an experimental signature through the production of photons + missing 
energy-momentum,
as is the case in the so-called  GMSB models.

\vspace{2mm}

Conversely, how the spin-$\frac{1}{2}$ Goldstone fermion (goldstino) field
couples to boson-fermion pairs determines how the boson and fermion masses get split within the multiplets of supersymmetry
\cite{grav,fayet79}.

\section{ELECTROWEAK BREAKING WITH TWO SPIN-0 DOUBLETS}

\label{sec:ewb}

\subsection{\boldmath Introducing $R$ symmetry, and $U(1)_A$, in a two-doublet model}

One of the initial difficulties in supersymmetric theories was to construct massive Dirac spinors carrying a
conserved quantum number that could be attributed to leptons, although supersymmetric theories
involve self-conjugate Majorana fermions which in principle cannot carry such a quantum number. This led to the
definition of $R$-symmetry, first obtained within a toy model for ``leptons'', soon reinterpreted as the charginos and
neutralinos of the Supersymmetric Standard Model \cite{R}.

\vspace{2mm}

This first supersymmetric electroweak model \cite{R} was obtained from a related pre-susy 2-spin-0-doublet vectorlike one \cite{2hd}, that was actually an ``inert doublet model'', 
close to being a supersymmetric theory. This one 
already included a $Q$ symmetry precursor of the $R$ symmetry,  acting on the two doublets ($\varphi"=h_1$ and $\varphi'=h_2^c$) according to
\be
\label{ua}
h_1\,\to \,e^{i\alpha}\,h_1\,,\ \ h_2\,\to \,e^{i\alpha}\,h_2\,.
\ee
This $Q$ symmetry, jointly with $U(1)_Y$, allowed to rotate independently the two doublets $h_1$ and $h_2$, 
restricting the structure of the Yukawa and quartic couplings very much as for Higgs and higgsino doublets within supersymmetry.
There it acts according to
\be
H_{1}\,(x,\theta)\ \stackrel{Q}{\rightarrow}\  e^{i\alpha} \ H_{1}\,(x,\theta\,e^{-\,i\,\alpha})\, ,\ \ \ 
H_{2}\,(x,\theta)\  \stackrel{Q}{\rightarrow}\  e^{i\alpha} \ H_{2}\,(x,\theta\,e^{-\,i\,\alpha})\, ,
\ee
allowing in particular for a $\mu\,H_1H_2$ mass term in the superpotential  (with the Higgs mass parameter $\mu$ equal to the higgsino one $m$).

\vspace{2mm}

This original definition of the $Q$-symmetry acting on the supersymmetry generator was then modified 
into the now-familiar definition of $R$-symmetry, acting according to
\be
\label{rhh}
H_{1}\,(x,\theta)\, \stackrel{R}{\rightarrow}\, H_{1}\,(x,\theta\,e^{-\,i\,\alpha})\, ,\ \ H_{2}\,(x,\theta)\, \stackrel{R}{\rightarrow}\, H_{2}\,(x,\theta\,e^{-\,i\,\alpha})\, ,
\ee
so as to leave $h_1$ and $h_2$ invariant 
and survive the electroweak breaking induced by $\,<h_1\!>\,$ and \hbox{$\,<h_2\!>$}\,, \,while forbidding a $\mu\,H_1H_2$ mass term in the superpotential.

\vspace{2mm}
Going from $Q$ to $R$ was done through the relation
\be
R\,= \,Q\,U^{-1}\,.
\ee
The additional $U(1)$ symmetry also defined in \cite{R}, later called $U(1)_A$, transforms $h_1$ and $h_2$ as in (\ref{ua}) 
but commutes with supersymmetry, in contrast with $Q$ and $R$. It acts according to
\be
\label{defua0}
H_1\ \stackrel{U(1)_A}{\rightarrow} \ e^{i\alpha}\,H_1,\ \ H_2\ \stackrel{U(1)_A}{\rightarrow} \ e^{i\alpha}\,H_2,
\ee 
also forbidding the $\mu H_1 H_2$ term in the superpotential (as what was called later a $U(1)_{PQ}$ symmetry),
with its definition extended to act axially on quark and lepton fields and superfields
\cite{ssm}. 
Just as $U(1)_A$ (and in contrast with $Q$), $R$ symmetry  forbids a $\mu\,H_1H_2$ mass term in the superpotential. This one was  
then replaced by an ``$R$-invariant'' trilinear coupling  $\,\lambda \,H_1 H_2 S$ with an extra singlet $S$ transforming according to
\be
\label{URS}
S(x,\theta)\  \stackrel{R}{\rightarrow} \ e^{2\,i\,\alpha}\ S(x,\theta\,e^{-\,i\,\alpha})\,,\ \ 
S\ \stackrel{U(1)_A}{\rightarrow} \ e^{-2i\alpha}\,S\,.
\ee
This continuous $R$-symmetry gets subsequently reduced to $R$-parity, in the presence of Majorana gravitino and gaugino mass terms.

\vspace{-1.5mm}

\subsection{Avoiding an ``axion''}

\vspace{-.5mm}

The $\mu$ term first considered in  \cite{R} does not allow to have both $\,<h_1\!>\,$ and $\,<h_2\!>$ non-zero, then leading to a massless chargino.
Indeed even in the presence of the weak-hypercharge $\xi D'$ term splitting the $h_1$ and $h_2$ mass$^2$ terms apart from $\mu^2$ we only get a non-vanishing 
v.e.v. for  one doublet, the other being ``inert''.
Taking $\mu=0$ to allow for $v_1$ and $v_2$ non-zero, with $\tan\beta=v_2/v_1$ (then denoted $\tan\delta= v'/v"$), would
only fix the difference $v_2^2-v_1^2$,  leaving us with {\it two flat directions} associated 
with the chiral superfield
\be
\label{HA0}
H_A=H_1^0 \,\sin\beta + H_2^0 \,\cos\beta \,,
\ee
 leading  to two classically-massless spin-0 bosons, $h_A$ and $A$.
 
 \vspace{2mm}
Indeed the field 
\be
\label{defA}
A=\sqrt 2\ \,\hbox{\rm Im} \, (\,h_1^0 \,\sin\beta + h_2^0 \,\cos\beta \,)\,,
\ee
orthogonal to the Goldstone combination 
$\,z_g=\sqrt 2\ \,\hbox{\rm Im}  \, (\,-\,h_1^0 \,\cos\beta + h_2^0 \,\sin\beta \,)$ eaten away by the $Z$, 
is associated with the breaking of the $U(1)_A$ symmetry in (\ref{defua0}). 
\vspace{-.2mm}
It
corresponds to an axionlike or even axion pseudoscalar once quarks get introduced, present in the mass spectrum \cite{R,ssm}
together with its corresponding real part, 
\be
\label{defha}
h_A=\sqrt 2\ \,\hbox{\rm Re} \, (\,h_1^0 \,\sin\beta + h_2^0 \,\cos\beta \,)\,,
\ee
  (``saxion'').
Both would remain classically massless in the absence of the extra singlet $S$, with superpotential interactions breaking explicitly the $U(1)_A$ symmetry
\footnote{With $\,\tan\beta=v_2/v_1\equiv\tan\delta=v'/v", \ h_1^\circ=\varphi"^\circ,\ h_2^0=\varphi'^{0*}$, 
\,the complex field $\,\varphi"^{0*} \sin\delta+\varphi'^0\cos\delta=(\,h_1^0 \,\sin\beta + h_2^0 \,\cos\beta \,)^*$, massless 
in the absence of $S$ \cite{R}, 
represents the would-be axionlike boson $A$ in (\ref{defA}) and associated scalar $h_A$.}. But there was no need then to explain 
how we got rid of such an unwanted spin-0 ``axion''  $A$ by breaking explicitly the $U(1)_A$ symmetry; indeed 
this notion was brought to attention three years later \cite{ww}, in connection with a possible solution to the $CP$ problem.

\vspace{2mm}
Avoiding such an ``axion'' and classically-massless associated scalar $h_A$ was done through 
an explicit breaking of the $U(1)_A$ symmetry (now often known as a PQ symmetry). It was realized through the superpotential interactions $f(S)$ of an extra singlet $S$ coupled 
through a  trilinear superpotential term $\lambda \,H_1 H_2 S$. This one is  invariant under $U(1)_A$,  defined in \cite{R} as
\be
\label{u1a}
H_1\ \stackrel{U(1)_A}{\rightarrow} \ e^{i\alpha}\,H_1,\ \ H_2\ \stackrel{U(1)_A}{\rightarrow} \ e^{i\alpha}\,H_2,\ \ \,S\ \stackrel{U(1)_A}{\rightarrow} \ e^{-2i\alpha}\,S\,.
\ee
The superpotential  interactions $f(S)$, including terms proportional to $S$, $S^2$ or $S^3$, break explicitly $U(1)_A$, 
so that a massless or light axionlike spin-0 boson $A$, in particular, is avoided. (This one may of course reappear with a small mass, in the presence of a small explicit 
breaking of $U(1)_A$, as e.g. in the $U(1)_A$ limit of the NMSSM, with a small $\frac{\kappa}{3}\, S^3$ superpotential term.)

\vspace{2mm}

Selecting among possible $f(S)$ interactions,
including $S$, $S^2$ or $S^3$ terms as in the general NMSSM,
the sole linear term proportional to $S$ presented the additional interest of leading to an ``$R$-invariant''  (nMSSM) superpotential \cite{R} 
 \be
 \label{nmssm}
{\cal W} \, = \, \lambda\,H_1 H_2\,S +\, \sigma\,S\,.
\ee
 Indeed $H_1,H_2$ and $S$ transform  as
\be
\label{trhs}
H_{1,2}\,(x,\theta)\  \stackrel{R}{\rightarrow}\ H_{1,2}\,(x,\theta\,e^{-\,i\,\alpha})\, ,\ \ \ 
S(x,\theta)\  \stackrel{R}{\rightarrow} \ e^{2\,i\,\alpha}\ S(x,\theta\,e^{-\,i\,\alpha})\,,
\ee
so that the superpotential (\ref{nmssm}) transforms with $R=2$ according to
 \be
 \label{wr0}
{\cal W}\,(x,\theta)\ \stackrel{R}{\rightarrow}\ e^{2\,i\,\alpha}\ {\cal W}\,(x,\theta\,e^{-\,i\,\alpha})\,,
\ee
as required for its last $F$ component to be $R$-invariant.

\vspace{2mm} 
We then get a conserved additive quantum number $R$ carried by the supersymmetry generator and associated with this unbroken $U(1)_R$ symmetry, 
in fact the progenitor of $R$-parity, $R_p= (-1)^R$. Such a superpotential has also the interest of triggering spontaneous electroweak breaking 
even in the absence of any supersymmetry breaking, in contrast with the MSSM.

\vspace{2mm}
Another possibility to avoid the would-be ``axion'' $A$  is to eliminate it by gauging the extra $U(1)_A$  (taking  $f(S)=0$ and assuming anomalies appropriately cancelled), as in the USSM \cite{ssm}.

\section{GAUGE/BE-HIGGS UNIFICATION WITHIN SUPERSYMMETRY}

\label{sec:gh}

\subsection{ \boldmath The massive gauge multiplet for the $Z$ boson}

In the first electroweak model \cite{R}
the Goldstone fermion was related to the photon by supersymmetry.
I.e.~the spin-$\frac{1}{2}$ Goldstone fermion, 
now called the goldstino, was identical to the spin-$\frac{1}{2}$
partner of the photon, known as the photino. Only charged particles are then sensitive to supersymmetry breaking, namely the $W^\pm$, 
the charged BE-Higgs boson $H^\pm$ (then called $w^\pm$) and their associated charginos.
The neutral ones, uncoupled to the photon and thus to the goldstino, still remain mass-degenerate implying at this stage \cite{R}
\be
\label{gh}
m_Z\,=\,m\,\hbox{(Dirac zino)} \,= \,m\,\hbox{(neutral spin-0 BEH boson)}\,.
\ee
The Dirac zino is obtained from chiral gaugino and higgsino components transforming in opposite ways 
under $R$ sym\-metry, according to
\be
\label{ghr}
\hbox{\em gaugino}\ \stackrel{R}{\rightarrow}\ e^{\gamma_5\alpha}\ \hbox{\em gaugino}\,,\ \ \ 
\hbox{\em higgsino}\ \stackrel{R}{\rightarrow}\ e^{-\gamma_5\alpha}\ \hbox{\em higgsino}\,.
\ee
 This $R$ quantum number, first presented as a toy-model ``lepton number'',
was reinterpreted as corresponding to a new class of $R$-odd particles, known as charginos and neutralinos.

\vspace{2mm}

In present notations,  $\,<\!h_i^0>\, =v_i/\sqrt 2\,$ with $\,\tan\beta=v_2/v_1$ 
replacing the original $\tan\delta=v'/v"=v_2/v_1$,  the Goldstone field eaten away by the  $Z$ is described by 
$\,z_g=\sqrt 2\ \,\hbox{\rm Im}  \, (\,-\,h_1^0 \,\cos\beta + h_2^0 \,\sin\beta \,)$, orthogonal to the pseudoscalar combination 
$A$ 
 in (\ref{defA}).
Together with the corresponding real part denoted by $z$ in \cite{R} \footnote{There we defined 
$z=\varphi"^0_1\cos\delta -\varphi'^0_{\ 1}\sin\delta=\, \sqrt 2\ \,\hbox{\rm Re} \, (\,h_1^0\,\cos\beta  - \,h_2^0\,\sin\beta \, ) $. 
We include here an optional change of sign in the definition (\ref{z}) of $z$, to avoid a $-$ sign for large $\tan\beta$.},
\be
\label{z}
z\,=\, \sqrt 2\ \,\hbox{\rm Re} \, (\,-\,h_1^0\,\cos\beta  + \,h_2^0\,\sin\beta \, )\,,
\ee
it is described by the chiral superfield combination,
\be
\label{hzzz}
H_z=-\,H_1^0\,\cos\beta  +\,H_2^0\,\sin\beta\,=\,\frac{z+iz_g}{\sqrt 2}+\,...\,,
\ee

\vspace{-3mm}
\noindent
orthogonal to 
\be
H_A=H_1^0 \,\sin\beta + H_2^0 \,\cos\beta \,=\,\frac{h_A+iA}{\sqrt 2} +\,...
\ee
describing the pseudoscalar $A$ as in (\ref{defA}).
This scalar $z$ is, independently of the value of $\beta$, the spin-0 field getting related with the $Z$ under supersymmetry,
with a mass term 
\be
\label{z2}
-\,\frac{1}{2}\ m_Z^2\,z^2
\ee
in the Lagrangian density. 
\vspace{2mm}

One of the first implications of supersymmetric theories is thus the possible existence, 
{\it independently of $\tan\beta$}, of a neutral spin-0 BE-Higgs boson
degenerated in mass with the $Z$, i.e. of mass (say $m_h$, for the reader used to MSSM notations)
\be
m_h \,\simeq \, 91\  \hbox{GeV}/c^2\ \ \hbox{\it up to supersymmetry-breaking effects}.
\ee
Neutral (or charged) {\it spin-0 BE-Higgs bosons get associated with  massive gauge bosons}, and related inos, 
within massive gauge multiplets of supersymmetry \cite{R}, according to
\be
\label{gh2}
Z\ \  \ \stackrel{SUSY}{\longleftrightarrow }\ \ \ \hbox{2 Majorana zinos} \ \ \ 
\stackrel{SUSY}{\longleftrightarrow }\ \ \ \hbox{neutral spin-0 BEH boson}\,.
\ee
The two Majorana zinos are obtained, in the usual formalism, from mixings of neutral gaugino and higgsino compo-
nents transforming under $R$ as in (\ref{ghr}). The continuous $R$ symmetry gets subsequently reduced to $R$-parity through
the effects to the $\mu$ term (directly included as in the MSSM, or possibly resurrected from a translation of $S$), and
direct gaugino mass terms ($m_{1/2}$) generated from supergravity or radiative corrections. These have $\Delta R = \pm 2$ and mix
neutralinos into four Majorana mass eigenstates, from the two Majorana zinos in (\ref{gh}), the photino and the neutral higgsino described
by $H_A$, as in the MSSM. There may also be more, as in the presence of additional N/nMSSM or USSM singlinos described by the
singlet $S$ or an extra-$U(1)$ gauge superfield.

\subsection{\boldmath Yukawa couplings ``of the wrong sign'' for  the $z$, spin-0 partner of the $Z$}

The new boson found at CERN with a mass close to 125 GeV/$c^2$
\cite{higgs,higgs2,hat,hcms}, believed to a BE-Higgs boson associated with the electroweak breaking, 
may well also be interpreted, in general up to a mixing angle as we shall see, as a spin-0 partner of the 91 GeV/$c^2$ $Z$ 
boson under {\it two} infinitesimal supersymmetry transformations.

\vspace{2mm}
We can compare the $z$ field in (\ref{z})
with the SM-like scalar field,
\be
\label{hsm}
h_{SM}=\, \sqrt 2\ \,\hbox{\rm Re} \, (\,h_1^0\,\cos\beta  + \,h_2^0\,\sin\beta \, )\,,
\ee
so that

\vspace{-7mm}

\be
\label{angle}
<\,h_{SM}\,|\ z>\ \ =\, -\,\cos 2\beta\,.
\ee
The two fields are at an angle $\,\pi -2\beta\,$ in field space, getting very close for large $\tan\beta$. 
The spin-0 $z$, directly related with the $Z$ under supersymmetry, tends for large $\tan\beta$ to behave very much as the SM-like 
$\,h_{SM}$.

\vspace{2mm}
Furthermore, while the SM-like scalar field $h_{SM}$ has Yukawa couplings to quarks and charged leptons 
\be
\frac{m_{q,l}}{v}\,= \,2^{1/4}\,G_F^{1/2}\ m_{q,l}\,,
\ee
the $z$ field, spin-0 partner of the spin-1 $Z$, has almost-identical Yukawa couplings 
\be
\label{zcoup}
\frac{m_{q,l}}{v}\ 2\,T_{3\,q,l}\,= \ 2^{1/4}\,G_F^{1/2}\ m_{q,l}\ 2\,T_{3\,q,l}\,,
\ee
simply differing by {\it a relative change of sign for $d$ quarks and charged leptons} (with $2\,T_{3\,d,l}=-1$) which acquire their masses through $<h_1^0>$, as compared to $u$ quarks. This may also be understood from the expression of the axial part in the weak neutral current $J_Z^\mu=J_3^\mu-\sin^2 \theta \,J^\mu_{\rm em}$, proportional to $J^\mu_{3\ \rm ax}$, the $Z$ boson coupled to $J_Z^\mu$ getting its mass by eliminating the would-be Goldstone boson $z_g$ that is the pseudoscalar partner of the spin-0 $z$, as seen from (\ref{hzzz}).

\vspace{2mm}

The $z$ has however {\it reduced trilinear couplings to the $W^\pm$ and $Z$}, by a factor $-\cos2\beta$, with

\vspace{-2mm}

\be
\left\{\ \ba{ccl}
(z\,VV)\   \hbox{couplings}\!&=&\! (h_{SM}\,VV)\ \hbox{couplings}\,\ \times\, \ \  (-\cos 2\beta)\,
\vspace{1mm}\\
(z\,ff)\   \hbox{couplings}\!&=&\! (h_{SM}\,VV)\ \hbox{couplings}\ \,\times\, \left( \,2T_{3f}\,=\left\{\ba{cl}  +1& u,c,t \vspace{.5mm}\\
-1& d,s,b;\ e,\mu,\tau
\ea\right\}\,\right)
\ea\right.
\ee
The expected production of a spin-0 $z$ in the $ZZ^{*}$  or $WW^{*}$ decay channels would then be {\it decreased by $\cos^2 2\beta$}
as compared to a SM boson, with respect to fermionic quark and lepton channels (the change of sign in $d$-quarks and charged-lepton couplings 
also affecting the $h\to\gamma\gamma$ decay).
\pagebreak

But  {\it  the $z$ field does not necessarily correspond to a mass eigenstate}, and 
further mixing effets induced by supersymmetry breaking must be taken into account, as discussed soon for the MSSM in Sec.\,\ref{subsec:mssm}.

\vspace{2mm}

Additional information on the production and decay rates of the new boson may tell 
whether it can originate from a single doublet as in the standard model, or if two doublets $h_1$ and $h_2$ 
are also allowed or possibly required.
The role of the
spin-0 combination $z$ in (\ref{z}) as related to the $Z$ by two supersymmetry transformations is then guaranteed
if supersymmetry is indeed relevant, even if no 
``supersymmetric particle'' has been found yet.

\vspace{2mm}

{\it The observation of a new spin-0 particle with a mass not too far from $m_Z$, possibly 
related to the $Z$ by supersymmetry, thus appears as an important indication in favor of this symmetry.}

\vspace{2mm}

According to this gauge-Higgs unification (already within $N=1$ theories in 4 dimensions), BEH bosons naturally appear as 
{\it extra spin-0 states of massive spin-1 gauge bosons}. And this, in spite of the fact  that they have {\it different gauge-symmetry properties} 
\,-- thanks to the spontaneous breaking of the electroweak symmetry. We also have, in a similar way,

\vspace{-4mm}
\be
\label{gh3bis}
W^\pm\ \ \ \stackrel{SUSY}{\longleftrightarrow }\ \ \ \hbox{2 Dirac winos} \ \  \ \stackrel{SUSY}{\longleftrightarrow }\ \ \ \hbox{charged  
spin-0 boson} \ H^\pm,
\ee
with a mass term 
\be
-\,\ m_W^2\,|H^\pm|^2\,,
\ee
up to supersymmetry-breaking effects. This is why the charged boson now known as $H^\pm$, appearing as a spin-0 partner of
the massive $W^\pm$,  was initially denoted $w^\pm$ in \cite{R}.

\subsection{\boldmath Charged and neutral spin-0 \,BE-Higgs bosons as described by $W^\pm$ and $Z$ massive gauge superfields}

Even more remarkably, these massive spin-1, spin-$\frac{1}{2}$ and spin-0 particles may all be described by  
(neutral or charged) {\it massive gauge superfields}  \cite{gh}. 
This is true  {\it in spite of their different electroweak gauge symmetry properties}, spin-1 fields 
transforming as a gauge triplet and a singlet while spin-0 BEH fields transform as electroweak doublets.
And although gauge and BE-Higgs bosons have {\it very different couplings to quarks and leptons}, 
BEH bosons being coupled proportionally to masses as seen from (\ref{zcoup}), in contrast with gauge bosons.
This may first appear very puzzling but is elucidated in \cite{gh}.
\vspace{2mm}

To do so we must {\it change picture} in our representation of such spin-0 bosons. The previous $z$ and $w^\pm$ ($\equiv H^\pm$) 
cease being described by spin-0 components of the chiral doublet BEH superfields $H_1$ and $H_2$, to get described,
through a non-polynomial change of fields, by the lowest ($C$) components of the $Z$ and  $W^\pm$ superfields, now massive.
This explicit association between massive gauge bosons and spin-0 BEH bosons 
can be realized in a manifestly supersymmetric way 
(at least for the $Z$ superfield for which supersymmetry stays unbroken at this stage) 
by completely gauging away the three (Goldstone) chiral superfields $H_1^-,\,H_2^+$ 
and $H_z = -\,H_1^0 \cos\beta+H_2^0 \sin\beta$, taken identical to their v.e.v.'s:
\be
\label{chirgold}
H_1^- \equiv H_2^+\equiv \,0\,,\ \ \ \ H_z = -\,H_1^0 \cos\beta+H_2^0 \sin\beta\,\equiv\,-\,\frac{v}{\sqrt 2}\,\cos 2\beta\ .
\ee

The corresponding $\,<\!H_1^0\!>\ =v_1/\sqrt 2\,$ and $\,<\!H_2^0\!>\ =v_2/\sqrt 2\,$  
\vspace{-.2mm}
generate mass terms 
$\,\frac{1}{2}\,m_Z^2 |Z|^2  $ and $\,m_W^2\, |W^+|^2 $  for the gauge superfields $Z(x,\theta,\bar\theta)$ and $W^\pm(x,\theta,\bar\theta)$.
The previous $z$ and $w^\pm$ ($\equiv H^\pm$) 
get described by the lowest ($C$) spin-0 components of these {\it massive} $Z$ and  $W^\pm $ superfields, expanded as
\be
Z(x,\theta,\bar\theta)\,=\, C_Z(x) + ...\, -\,\theta\sigma_\mu \bar\theta\   Z^\mu(x) + ...\, , \ \ \ W^\pm(x,\theta,\bar\theta)\,=\, C_W^\pm(x) + ...\, 
-\,\theta\sigma_\mu \bar\theta\      W^{\mu\,\pm }(x) + ...\, .
\ee
Their last (``$C$'') components now correspond to physical dynamical degrees of freedom describing,
through {\it non-polynomial field transformations},
linearized as
\be
z\,= \,m_Z \,C_Z+\,...\ ,\ \ \ w^\pm = \,m_W\,C_{\pm}+\,...\ ,
\ee
 the spin-0 BE-Higgs fields previously referred to as $z$ and 
$w^\pm$, now known as $H^\pm$ \cite{gh}:
\be
\hbox{\it massive gauge superfields now describe also spin-0 BEH fields !}
\ee
Their subcanonical (``$\chi$'')  spin-$\frac{1}{2}$ components 
\vspace{-.5mm}
also correspond to physical degrees of freedom, 
describing (again through non-polynomial field transformations) the spin-$\frac{1}{2}$ fields previously known as higgsinos.

\vspace{2mm}

We then keep only 
\be
\label{HA}
H_A= H_1^0 \sin\beta+H_2^0 \cos\beta =\,\frac{h_A+iA}{\sqrt 2}+\,...\,,
\ee
as an  (``uneaten'')  chiral superfield, describing as in (\ref{defA},\ref{defha})
the pseudoscalar 
$A=\sqrt 2\ \,\hbox{\rm Im}  \, (\,h_1^0 \,\sin\beta + h_2^0 \,\cos\beta \,)$ and associated scalar
$h_A=\sqrt 2\ \,\hbox{\rm Re} \, (\,h_1^0 \,\sin\beta + h_2^0 \,\cos\beta \,)$.

\subsection{\boldmath The \,BE-Higgs boson as spin-0 partner of $\,Z$, \,in the MSSM and beyond}
\label{subsec:mssm}

This applies in particular to the specific model known as the MSSM, here expressed as including 
the soft dimension-2 supersymmetry-breaking gauge-invariant mass term
\be
\label{ma}
-\,m_A^2\ \,| h_1 \sin\beta -\,h_2^c \cos\beta|^2\,.
\ee
It may be considered as an ``inert doublet'' mass term,
chosen to vanish for  $\, <h_i^0>\ =v_i/\sqrt 2$ \,with $\,\tan\beta=v_2/v_1$.
This term thus {\it does not modify the vacuum state considered}, initially taken as having a spontaneously-broken supersymmetry
in the gauge-and-Higgs sector, with the photino playing the role of the Goldstone fermion
so that the mass equalities (\ref{gh}) applies for neutral particles, with $m^2_{H^\pm}= m_W^2$ for charged ones \cite{R,ssm}.

\vspace{2mm}

The gauge-and-Higgs sector, first considered with a spontaneously-broken supersymmetry, admits 
two neutral classically-flat directions of the potential, 
associated with the scalar and pseudoscalar fields $h_A$ and $A$ described by the initially-massless chiral superfield
$H_A = H_1^0 \sin\beta+H_2^0 \cos\beta$ in (\ref{HA0},\ref{HA}).
There is indeed initially, in the absence of the inert-doublet mass term (\ref{ma}), a classically-massless pseudoscalar 
\be
\label{A}
A=\sqrt 2\ \,\hbox{\rm Im}  \, (\,h_1^0 \,\sin\beta + h_2^0 \,\cos\beta \,)\,,
\ee
associated with the breaking of the $U(1)_A$ symmetry 
$
h_1\,\to \,e^{i\alpha}\,h_1, \ h_2\,\to \,e^{i\alpha}\,h_2$ in  (\ref{ua},\ref{u1a}) \cite{R}, 
which acquires a mass $m_A$  from the added soft supersymmetry-breaking terms (\ref{ma}) 
that break explicitly $U(1)_A$.

\vspace{2mm}

Defining

\vspace{-6mm}
\be
\label{phiphi}
\left\{ \ba{ccc}
\varphi_{\rm sm}\!&=&\! h_1 \cos\beta +\,h_2^c \sin\beta\,,
\vspace{2mm}\\
\varphi_{\rm in}\!&=&\! h_1 \sin\beta -\,h_2^c \cos\beta\,,
\ea\right.
\ee
the two previously-flat directions associated with $A$ and $h_A$ get lifted by the mass term (\ref{ma}) for the inert
combination $\varphi_{\rm in}$.
With 
\be
\label{ma2}
|\varphi_{\rm in}|^2 = |\, h_1 \sin\beta -\,h_2^c \cos\beta\,|^2 =  \,
 |H^+|^2 + \frac{1}{2}\,A^2 + 
 \frac{1}{2}\ \,|\,\sqrt 2 \ \,\hbox{\rm Re} \,( h_1^0 \sin\beta -\,h_2^0 \cos\beta)\,|^2
\ee
it also provides an additional supersymmetry-breaking contribution $m_A^2$ to $m^2_{H^\pm}$
 in addition to the supersymmetric one $m_W^2$,  so that
 \vspace{-6mm}
 
\be
m^2_{H^\pm}= m_W^2 + m_A^2\,.
\ee
Adding the supersymmetric contribution $m_Z^2$ from (\ref{z},\ref{z2}) 
and supersymmetry-breaking one $m_A^2$ from (\ref{ma},\ref{ma2}) we get
{\it directly} \,the $2\times 2$ spin-0 scalar mass$^2$ matrix
\be
{\cal M}_\circ^2\,=\,\left(\ba{cc}
c_\beta^2\,m_Z^2+ s_\beta^2\,m_A^2 &-\,s_\beta c_\beta \, (m_Z^2+m_A^2)   \vspace{2mm}\\
- \,s_\beta c_\beta\,  (m_Z^2+m_A^2)   & s_\beta^2\,m_Z^2+ c_\beta^2\,m_A^2 
\ea\right)\,,
\ee
verifying
\be
\hbox{Tr}\,{\cal M}_\circ^2\,=\, m_H^2+m_h^2 = m_Z^2 + m_A^2\,,   
\ \ \ \ \,\hbox{det}\,{\cal M}_\circ^2\,=\, m_H^2 \,m_h^2 = m_Z^2 \,m_A^2\,\cos^2\,2\beta\,,
\ee
so that

\vspace{-7mm}
\be
m^2_{H,h}\,=\, \hbox{\small $\dis\frac{m_Z^2+m_A^2}{2}$}\pm 
\sqrt {\hbox{ \footnotesize$\dis\left(\frac{m_Z^2+m_A^2}{2}\right)^2 $}  -  m_Z^2 m_A^2\,\cos^2\,2\beta }\ 
 \ \ \ \ \hbox{\small ( + radiative corrections)}\,.
\ee

This implies  the well-known relation $m_h < m_Z |\cos 2\beta |$ at the classical level, up to radiative corrections 
which must be significant  if one is to reach  $\simeq 125$ GeV/$c^2$
from a classical value between 0 and $m_Z$. This classical value of $m_h$  can approach $m_Z$ for large $\tan\beta$ 
with large $m_A$ \,i.e.~large supersymmetry-breaking effects. We must then also count on signifiant quantum corrections 
from large supersymmetry breaking 
(e.g. from very heavy stop squarks), if the resulting  $m_h$
is to reach the observed $\,\simeq$ 125 GeV$/c^2$.

\vspace{2mm}
These scalar mass eigenstates behave for large $m_A$ as
\be
\left\{\ba{ccrcl}
H &\to& \ \ \sqrt 2 \ \,\hbox{\rm Re} \,( \, h_1^0 \sin\beta -\,h_2^0 \cos\beta)&& \ \hbox{with large mass}  \ m_H\simeq m_A\,,
\vspace{2mm}\\
h&\to&h_{SM}= \sqrt 2 \ \,\hbox{\rm Re} \,( h_1^0 \cos\beta +\,h_2^0 \sin\beta)&&\  \hbox{with standard-model couplings}\,.
\ea\right.
\ee
Indeed $\varphi_{\rm sm}= h_1 \cos\beta +\,h_2^c \sin\beta$ in (\ref{phiphi})
is the ``active'' SM-like Higgs doublet acquiring a v.e.v.,
with  $h_{SM}$ coupled as in the standard model.  And the $h$ field, presumably to be associated 
with the 125 GeV$/c^2$ boson observed at CERN, is very close to the $z$ in (\ref{z})
for large $\tan\beta$, then justifying {\it an almost  complete association of this 125 GeV$/c^2$ boson with the spin-1 $Z$}, as indicated by (\ref{angle}).

\vspace{2mm}
But we do not want to stick too closely to the specific case of the MSSM, as we felt 
from the beginning that its 2-doublet structure ought to be extended to the extra singlet $S$
with a trilinear superpotential coupling $\,\lambda\,H_1 H_2 S$. And as it is now very strongly constrained,
in many of its interesting regions in parameter space.
Furthermore this extra N/nMSSM singlet $S$ introduced by turning the $\mu$ parameter into a dynamical
superfield in superspace according to

\vspace{-6mm}

\be
\mu\ \ \to\ \ \mu(x,\theta)=\lambda\,S(x,\theta)\,,
\ee
leads to a new quartic spin-0 coupling independent of the gauge couplings, which helps making the lightest 
BE-Higgs boson sufficiently heavy.
Indeed  {\it the lightest neutral spin-0 mass may already be equal to $m_Z$ at the classical level,
even before any breaking of the supersymmetry}, and independently of $\tan\beta$~\cite{R}. This is in sharp contrast with the MSSM 
for which it would at best vanish (for $\mu=0$) or worse, in which one does not even get any electroweak breaking in the absence of
supersymmetry breaking (for $\mu \neq 0$).

\vspace{2mm}

But let us now turn to another direction explored in a parallel way, 
leading us from $N=1$ supersymmetric theories with a continuous $U(1)_R$ symmetry 
to $N=2$ and $N=4$ theories, naturally expressed in an 
extended spacetime with extra compact dimensions, before coming back to the $N=1$ Supersymmetric Standard Model
in Sec.\,\ref{sec:ssm}.

\section{\boldmath $\!$From $R$-symmetry to $\,N=2\,$ and $\,N=4\,$ supersymmetry, \,and extra dimensions}

\subsection{\boldmath $F$-breaking of supersymmetry, with $R$-symmetry}

Let us return to the second classical mechanism of  spontaneous breaking of the global supersymmetry, relying on non-vanishing v.e.v.'s for auxiliary $F$-components of chiral superfields \cite{F1,or}.
 In order to do so  {\it the superpotential must be very carefully chosen} to avoid the existence of one or usually several supersymmetric vacuum states with vanishing energy,
that would necessarily be stable, with supersymmetry remaining conserved. 
\vspace{2mm}

Indeed for $n$ interacting chiral superfields, the set of  $n$  analytic equations $F_i(\varphi_j) =0$ 
(of degree 2 at most for a renormalizable theory) for $n$ complex variables $\varphi_j$ has solutions
in almost all situations, excepted 
very special ones. All auxiliary $F$ components can then vanish simultaneously, leading to a stable vacuum state (or usually several)  for which supersymmetry is preserved, with $\,<\!H\!>\ = 0$.

\vspace{2mm}

This  choice of suitable superpotentials, for which such {\it supersymmetric vacua are totally avoided} (rather than just been made unstable,
which is not possible here as discussed in Sec.~\ref{sub:spec}) is realized with the help of an additional $R$ symmetry \cite{R} acting chirally on the supersymmetry generator according to
\be
\label{R}
Q\ \stackrel{R}{\rightarrow}\ e^{-\gamma_5\alpha}\ Q \,.
\ee
It relies both on $R=2$ and $R=0$ chiral superfields transforming according to
\be
\Phi (x,\theta) \ \  \stackrel{R}{\rightarrow}\ \ e^{i\,R_\Phi\,\alpha}\ \Phi(x,\theta\,e^{-\,i\,\alpha})\,,
\ee
products of superfields being allowed in the superpotential only if they verify $\sum R_\Phi=2$.
\vspace{2mm}

The original example of \cite{F1} involves, 
in nMSSM-like notations, two chiral doublets $H_1$ and $H_2$ having $R=0$, interacting with a triplet {\boldmath $T$}
 and a singlet $S$ having $R=2$, through the superpotential
\be
\label{wsup}
{\cal W} \,=\, H_1\, (\lambda \, \hbox{\boldmath $\tau$} .\hbox{\boldmath $ T$}
 +\lambda' S)\,H_2 + \sigma S\,,
\ee
with a global $SU(2)\times U(1)$ electroweak-like symmetry.
It is a global version of the electroweak model \cite{R} with the gauge superfields omitted, supplemented with an additional triplet
{\boldmath $T$} next to the (N/nMSSM) singlet $S$, already having in mind for its gauged version an $N=2$
extended supersymmetric, or ``hypersymmetric'' theory \cite{hyper}, with $(H_1,H_2)$ describing an $N=2$ (matter)
{\it hypermultiplet}. 
\vspace{2mm}

We thus have in this model of spontaneous supersymmetry breaking through auxiliary $F$-component v.e.v.'s  {\it the same number of $R=0$ and  $R=2$ chiral superfields}, 4 in each case. (This is also in connection with the under\-lying $N=2$ structure of the model 
when it is gauged,  and its extra global $SU(2)$ symmetry acting on the $N=2$ supersymmetry generators, softly broken through the weak-hypercharge $\xi D'$ 
and/or $\sigma F_S$ terms.) \ They transform under $R$ as in (\ref{rhh},\ref{URS}), according to
\be
\label{defr0}
\left\{
\ba{ccc} \ \ \ \ 
H_{1,2}\,(x,\theta)\ \stackrel{R}{\rightarrow}\ H_{1,2}\,(x,\theta\,e^{-\,i\,\alpha})\,, 
\vspace{3mm}\\
S(x,\theta)\ \stackrel{R}{\rightarrow}\ e^{2\,i\,\alpha}\ S(x,\theta\,e^{-\,i\,\alpha})\,,  \ \ \   
\hbox{\boldmath $T$}(x,\theta)\ \stackrel{R}{\rightarrow}\ e^{2\,i\,\alpha}\ \hbox{\boldmath $ T$}(x,\theta\,e^{-\,i\,\alpha})\,.
\ea
\right.
\ee
$R$-symmetry requires that the superpotential ${\cal W}$ be {\it a linear function of the $R=2$ superfields}, 
$S$ and {\boldmath $T$}, the triplet {\boldmath $T$} being excluded by the $SU(2)$ electroweak-like symmetry.
It  excludes in particular a $\mu\, H_1 H_2$ superpotential term, 
so that ${\cal W}$ transforms with $R=2$ as in (\ref{wr0}) \cite{R,F1}.

\vspace{2mm}
With such suitably-chosen superpotentials  \cite{F1,or}  all auxiliary $F$-components cannot vanish simultaneously, the set of equations $F_i(\varphi_j) =0$ 
being constructed so as to have no solution, thanks in particular to the requirement of $R$-symmetry.
As a result supersymmetry gets spontaneously broken.
 A systematic consequence of this mechanism is the existence of an infinite set of inequivalent classically-degenerate vacua
associated with {\it classically-flat directions} (valleys) corresponding to classically-massless particles 
({\it pseudomoduli}) other than Goldstone bosons.

\vspace{2mm}

This includes in particular two classically-flat directions associated with the spin-0 component of one of the $R\!=\!2$ 
chiral superfields, leading to the possibility of discussing, depending on radiative corrections, the spontaneous (or quasi-spontaneous) breaking of $R$-symmetry. 
It would  lead to a massless $R$ Goldstone boson, if the $R$ symmetry is non-anomalous; or to an $R$-axion, as the $R$ symmetry, 
which acts axially on gluinos, is usually anomalous \cite{U}.

\subsection{\boldmath From $N\!=\!1$ with a $U(1)_R$ symmetry to $N\!=\!2$}

We now consider  the global $SU(2)\times U(1)$ symmetry of this model of interacting chiral superfields 
\cite{F1} as returning to local,
as in \cite{R} extended by an extra chiral triplet {\boldmath $ T$} next to the nMSSM singlet $S$.  
For a special choice of the trilinear superpotential couplings $\lambda$ and $\lambda'$ in (\ref{wsup}) in terms of the gauge couplings 
$g$ and $g'$ the gaugino fields  ($\lambda_L$) described by the electroweak gauge superfields may be related to the 
spin-$\frac{1}{2}$ fermion fields ($\zeta_L$)
described to the singlet and triplet chiral superfields $S$ and {\boldmath $T$} 
through an extra global $SU(2)$ symmetry, also acting on the two spin-0 doublets ($\varphi"=h_1$ and $\varphi'=h_2^c$) described by  $H_1$ and $H_2$
but not on their higgsino counterparts. It thus acts on the $N=2$ supersymmetry generators themselves, now transforming as a chiral doublet 
under this $SU(2)$ symmetry \cite{hyper}.

\vspace{2mm}

Then 

 \vspace{-7mm}
 
\be
\label{gino}
\left(\ba{c}  \lambda_L \vspace{.5mm} \\  \zeta_L \ea\right)\ \to\ \ SU(2)\hbox{-doublets of} \ \ N=2 \ \ \hbox{\em gauginos} 
\ee
transform as  (global) $SU(2)$ doublets of left-handed spinors with $R=1$.
This places the fermions $\zeta_L$ in adjoint chiral multiplets on the same footing as the adjoint chiral $\lambda_L$
associated with the Majorana gauginos in the gauge multiplets, upgrading $\zeta_L$ up to a new gaugino status. 
$ \lambda_L$ and $\zeta_L$ in (\ref{gino}) globally behave as an isodoublet of gauginos, for an enlarged $N=2$ 
extended supersymmetry (``hypersymmetry'') algebra. 
The two chiral doublets $H_1$ and $H_2$ responsible for electroweak breaking jointly describe 
electroweak doublets of spin-$\frac{1}{2}$ and spin-0 fields forming  an $N=2$ {\it ``hypermultiplet''}. 
Each hypermultiplet describes a Dirac spinor and two complex spin-0 fields, 
i.e. 4 fermionic + 4 bosonic field degrees of freedom.

\vspace{2mm}

This additional $SU(2)$ symmetry leading to $N=2$ supersymmetry 
requires 
$H_1$ and $H_2$ to interact with trilinear superpotential couplings as in (\ref{wsup}) but fixed by $g$ and $g'$ according to
$\lambda =g/\sqrt 2,\ \lambda' =g'/\sqrt 2$, so that
\be
\label{wsup2}
{\cal W} \,=\, \hbox{\small $\dis \frac{1}{\sqrt 2}$}\ H_1\, (g\,\hbox{\boldmath $\tau$} .\hbox{\boldmath $ T$} 
+g' S)\,H_2 + \sigma S\,,
\ee
the trilinear superpotential terms getting totally fixed by the gauge couplings.

\pagebreak

$D$-breaking and $F$-breaking  mechanisms for $N=1$ spontaneous supersymmetry breaking then get unified within $N=2$.
It allows for an abelian $\xi D'$ term responsible for $D$-breaking within a non-abelian theory \cite{R}
to be rotated into a related $\sigma F_S$ term responsible for $F$-breaking \cite{F1}, through a global $SU(2)$ rotation 
acting on the $N=2$ supersymmetry generators \cite{hyper}. This theory involves triplet and singlet chiral superfields 
{\boldmath $T$} and $S$, next to the $H_1$ and $H_2$ doublets, with a superpotential first restricted by $R$-invariance
as in (\ref{wsup}), then by the (softly-broken) global $SU(2)$ as in (\ref{wsup2}).
In this specific example the two Goldstone fermions (goldstinos) are the two photinos, related to the photon and two additional 
{\it spin-0 photons} by $N=2$ supersymmetry. Only charged particles are then sensitive at lowest order to the spontaneous breaking of the supersymmetry.
Neutral ones remain mass degenerate within two $N=2$ gauge multiplets, a massless one with the photon  and a massive one associated with  the $Z$.
\vspace{2mm}

Indeed the two neutral ($N=1$) gauge superfields associated with $W_3$ and $W'$ (i.e.~the $Z$ and photon superfields)
join the four neutral chiral ones $T_3,\,S,\,H^0_1$ and $H^0_2$ to describe, ultimately, a massive
$N=2$ gauge multiplet (with the $Z$, 4 Majorana zinos and 5 spin-0 bosons), and a massless one (with the
photon, 2 Majorana photinos and 2 spin-0 photons).  
This leads us to discuss again gauge/BE-Higgs unification, 
this time within $N=2$ theories.

\subsection{\boldmath Gauge\,/\,BE-Higgs unification in $N=2$}

As we just saw the $N=2$ associations 
\be
\label{gh40}
\gamma\ \  \ \stackrel{N=2}{\longleftrightarrow }\ \ \ \hbox{\it 2 Majorana photinos} \ \ \ \stackrel{N=2}{\longleftrightarrow }\ \ \ \hbox{\it 2 neutral spin-0 photons}\,,
\ee
and
\vspace{-4mm}
\be
\label{gh4}
Z\ \  \ \stackrel{N=2}{\longleftrightarrow }\ \ \ \hbox{\it 4 Majorana zinos} \ \ \ \stackrel{N=2}{\longleftrightarrow }\ \ \ \hbox{\it 5 neutral spin-0 bosons}\,,
\ee
were initially obtained in the $N=2$ (hypersymmetric) $SU(2) \times U(1)$ elecroweak model \cite{hyper},
with  the $N=2$ supersymmetry spontaneously broken through one, or a combination,
of the $D$-breaking \cite {R,fi} and $F$-breaking \cite{F1} mechanisms, becoming equivalent and unified within $N=2$. The two associated 
goldstinos are then the two gaugino partners of the photon, known as photinos, within a $N=2$ gauge multiplet. 
As for $N=1$ in \cite{R} boson-fermion mass splittings are felt only by charged particles. 
Neutral ones remain mass-degenerate at the classical level within the massive $Z$ and massless photon multiplets of $N=2$ supersymmetry, as shown above in (\ref{gh40},{\ref{gh4}).

\vspace{2mm}
Extending (\ref{gh40}) to QCD leads to
\be
\label{gh41}
gluons\ \  \ \stackrel{N=2}{\longleftrightarrow }\ \ \ \hbox{\it 2 Majorana gluino octets} \ \ \ \stackrel{N=2}{\longleftrightarrow }\ \ \ 
\hbox{\it 2 neutral spin-0 gluon octets}\,.
\ee
Extending (\ref{gh4}) to the $W^\pm$ requires {\it four doublet BE-Higgs superfields} $H_1,H'_1$ and $H_2,H'_2\,$ rather than the usual two, so as describe, altogether, a massive $N=2$ \,gauge multiplet said to be {\it of type I} \,as for the $Z$ multiplet in (\ref{gh4}) (cf. Sec.\,\ref{subsec:type} soon), with
\be
\label{gh5}
W^\pm\ \  \ \stackrel{N=2}{\longleftrightarrow }\ \ \ \hbox{\it 4 Dirac winos} \ \ \ \stackrel{N=2}{\longleftrightarrow }\ \ \ \hbox{\it 5 charged spin-0 bosons}\,.
\ee

\vspace{2mm}

This attractive property of gauge-Higgs unification  applies to other gauge bosons including those 
associated with grand-unified theories, or with an extra-$U(1)$ gauge group. 
These theories may also be obtained from an extended space-time with additional compact dimensions \cite{24,56,56bis}, with 
\be
\label{vhat}
\hbox{\em spin-0 photons, gluons, ... \ \ $\leftrightarrow$ \ \ extra components of  6d gauge fields}\ \  
V^{\hat\mu}=\left(\ba{l} V^\mu\vspace{.5mm}\\ V^5=a\vspace{.5mm}\\ V^6=b\ea\right)\,,
\ee
etc..
The latter associations correspond to {\it a different type of gauge/BE-Higgs unification} as compared to the one discussed earlier for $N=1$ 
supersymmetry in Sec.\,\ref{sec:gh} \cite{R,gh}. This one was conceptually more subtle by relating spin-1 and spin-0 fields 
with different gauge symmetry properties, in contrast  with (\ref{vhat}). Both types are physically relevant and get combined when dealing with supersymmetric GUTs with extra dimensions \cite{56,56bis}.
This also corresponds, in 4 dimensions, to different types of massive gauge multiplets of $N=2$, of types {\it I}, {\it II} or {\it III}, as will be discussed soon  in Sec.\,\ref{subsec:extra}.

\vspace{2mm}
This opens the question of the breaking of the $N=2$ supersymmetry, even harder than for $N=1$
especially if we also aim at a realistic theory with quarks and leptons acquiring masses 
from Yukawa couplings to spin-0 doublets,  and without abandoning too much of the extended symmetries 
that the theory is supposed to have.
Before that, let us pursue for a while in the direction of further increasing the symmetry.

\subsection{\boldmath From $N\!=\!2$ to $N\!=\!4$}

 A further step in this direction of increasing the symmetry is obtained with a $N=2$ supersymmetric Yang-Mills theory 
 describing a $N=2$ gauge multiplet interacting with a massless  spin-$\frac{1}{2}$-spin-0 hypermultiplet 
 in the adjoint representation of the gauge group. This leads to \cite{24}
\be
\left\{\ba{c}
N=2\ \ \hbox{\it supersymmetric Y-M theory} \  
\vspace{2mm}\\
\hbox{\small with}\ \ N=2\  \ \hbox{\it adjoint ``matter'' hypermultiplet}
\ea\right.\ \ \ \to\ \ \ N=4\ \ \hbox{\it supersymmetric Y-M theory} \,,
\ee
involving a set of four chiral adjoint gauginos 
\be
\left(\ba{c}  \lambda_{L} \vspace{.5mm} \\  \zeta_{1L} \vspace{.5mm} \\  \zeta_{2L} \vspace{.5mm} \\  \zeta_{3L} \ea\right)
\ \to\ \ SU(4)\hbox{-quartet of} \ \ N=4 \ \ \hbox{\em gauginos.} 
\ee
They transform as a quartet of the global  $\,SU(4)$ \,($\sim O(6)$) symmetry group acting on the set of $N=4$ supersymmetry generators, 
also transforming as a chiral quartet of $SU(4)$. 
These theories describe, from the three chiral adjoint $N=1$ superfields involved, a $SU(4)$ sextet of spin-0 fields in the adjoint representation of the gauge group, so that
\be
N=4 \ \ \hbox{gauge multiplet}   \ = \  (1 \ \hbox{spin-1}  \, + \,4 \  \hbox{spin-\small$\dis \frac{1}{2}$} + 6 \ \hbox{spin-0})\ \  \hbox{\it adjoint gauge fields}\,.
\ee

 These $N=4$ supersymmetric Yang-Mills theories  may be obtained directly from $N=2$ theories in 4 dimensions \cite{24}, 
or equivalently  from dimensional reduction of a $N=1$ supersymmetric Yang-Mills theories in 10 dimensions \cite{n4bis}.
They involve a single gauge coupling with no arbitrary Yukawa or quartic couplings, and  are totally fixed in 4 dimensions up 
to the choice of the Yang-Mills group  \,-- and vacuum state for which gauge symmetry may be spontaneously broken. This implies, however, a reduced flexibility
taking us farther away from a realistic theory of particles and interactions.

\subsection{\boldmath Spontaneous generation of central charges in $N=2$ or $4$ supersymmetry algebras}

A remarkable feature of such $N=2$ and $N=4$  theories is that the supersymmetry algebra gets {\it spontaneously modified} when the Yang-Mills symmetry 
gets spontaneously broken \cite{24}. Indeed an intriguing phenomenon occurs, 
which had to be properly elucidated before asserting that we are actually dealing with a {\it bona fide} $N=4$ supersymmetry algebra,
supposed to be
\be
\label{alg2}
\left\{ \  
\begin{array}{ccc}
\{ \ Q_i\, , \, \bar Q_j \ \} \!&=&\! 
- \, 2\,\gamma_{\mu}   P^{\mu} \, \delta_{ij}\,,\vspace {2mm} \cr 
[ \ Q_i\,, \, P^{\mu} \,] \!&=& \ 0\ .
\end{array}  \right.      
\ee
But,
when the Yang-Mills symmetry gets spontaneously broken, with the $N=4$ supersymmetry generators $Q_i$ remaining unbroken, we generate 
{\it a new sort of massive multiplet}, necessarily complex. Each one describes  1 spin-1, 4 Dirac spin-$\frac{1}{2}$ and 5 spin-0 fields,
with (6+10) bosonic + 16 fermionic degrees of freedom altogether  \cite{24}.  
This corresponds to a massive multiplet of particles with maximum spin 1, which however  is {\it not a representation of the $N=4$ supersymmetry algebra} (\ref{alg2})\,! 
\,So, what is going on\,?

 \vspace{2mm}

Actually the supersymmetry algebra (\ref{alg2}) is valid up to field-dependent gauge-transformations 
(and terms proportional to field equations of motion).
When the Yang-Mills symmetry gets spontaneously broken through the translation of some of the adjoint spin-0 gauge fields, 
these field-dependent gauge transformations acquire spontaneously-generated constant parts.
They correspond to some of the unbroken Yang-Mills generators, now {\it promoted to abelian} as the other Yang-Mills generators, with which they would not commute, get spontaneously broken.  They thus belong  to the center of the (super)symmetry algebra.

 \vspace{2mm}
These {\it spontaneously-generated central charges} then appear in the right-hand side of the anticommutation relations (\ref{alg2}) \cite{24}.
This extended supersymmetry algebra gets thus spontaneously-modified,
to include central charges in its anticommutation relations.
This leads to the same kind of algebra as in \cite{hls}, even if its conclusions cannot be applied directly 
as its conditions of validity are not met.

 \vspace{2mm}
 As we shall see these central charges play an essential role in the framework of grand-unification \cite{gut,gut2}, 
 when moving from the standard model to a grand-unification gauge group
 like $SU(5)$ or $O(10)$, ...\,.

\subsection{\boldmath Massive multiplets for $N=2$ grand-unification with gauge/BE-Higgs unification}
 \label{subsec:type}
 Modifying the algebra, spontaneously or not, to include central charges in the right-hand side of the anticommutation relations (\ref{alg2})
 allows for new massive multiplets (sometimes referred to as BPS) to appear as representations of this algebra.
 We get in particular new massive (``short'') multiplets with maximum spin $N/4$, instead of $N/2$ in the absence of such central charges.
 The first example is {\it the massive (matter) hypermultiplet of $N=2$}, describing a Dirac spin-$\frac{1}{2}$ and 2 spin-0 particles, all charged, 
 with 4 bosonic + 4 fermionic degrees of freedom \cite{hyper}.
 Another example is {\it the massive gauge multiplet of $N=4$}, describing 1 spin-1, 4 Dirac spin-$\frac{1}{2}$ and 5 spin-0 particles, charged, with 16 + 16 \,d.o.f. altogether \cite{24}.

 \vspace{2mm}
 The $N=2$  massive gauge multiplet such as the $Z$ one in (\ref{gh4}), real,  does not admit a central charge.
 It describes 
 1 spin-1, 2 Dirac spin-$\frac{1}{2}$ (or 4 Majorana) fermions and 5 spin-0 particles, with 8 + 8 \,d.o.f..
 And similarly for the $W^\pm$ multiplet in (\ref{gh5}), charged,
 with 16 + 16 \,d.o.f..  These massive gauge multiplets {\it of type I}, relevant for the description of electroweak interactions, 
 do not require a central charge ($Z$) in the anticommutation relations, even though they may be complex as for the $W^\pm$ multiplet.

 \vspace{2mm}

  Other types of $N=2$ massive gauge multiplets, however, require a central charge and are necessarily complex.
  They play a crucial role in $N=2$ extended supersymmetric grand-unified theories \cite{56bis}.
 The massive gauge multiplet {\it of type II\,} describes  1 spin-1, 2 Dirac spin-$\frac{1}{2}$ and 1 spin-0 particles, all charged,
with 8 + 8 \,d.o.f., times 3  when dealing with $SU(3)$ antitriplets or triplets.
This ``smaller'' multiplet is relevant  to describe the $X^{\pm 4/3}$ in a $SU(5),\ O(10)$ or $E(6)$ ... grand-unified theory, 
with a central charge $Z=\pm\,m_X$ in the right-hand side of the anticommutation relations of the $N=2$ supersymmetry generators.

 \vspace{2mm}
 
  Yet another type of multiplet, {\it of type III}, has the same field field content as a complex {\it type I\,} multiplet, but with
  a non-vanishing value of the central charge $Z$.
  It describes 
 1 spin-1, 4 Dirac spin-$\frac{1}{2}$ fermions and 5 spin-0 particles, with 16 + 16 \,d.o.f. as for the $W^\pm$ multiplet, times 3  when dealing with $SU(3)$ antitriplets or triplets.
 \vspace{-.2mm}
 It is relevant  to describe the $Y^{\pm 1/3}$ in a grand-unified theory, with a central charge $Z=\pm\,m_X$, and a mass $\,m_Y=  \sqrt{m_X^2+m_W^2} > |Z| = m_X$.

 \vspace{2mm}

One ultimately gets, in a
$N=2$ $SU(5)$-type supersymmetric GUT, the following $X$ and $Y$ multiplets \cite{56bis}
\be
\label{ghxy}
\left\{\ 
\ba{ccccccc}
X^{\pm 4/3}\ \ & \stackrel{N=2}{\longleftrightarrow }&\ \  \hbox{\it 2 Dirac xinos}\ \  & \stackrel{N=2}{\longleftrightarrow }
&\ \  \hbox{\it 1 charged spin-0 boson}\ && \hbox{\it (type II)}
\vspace{2mm}\\
Y^{\pm 1/3}\ \ & \stackrel{N=2}{\longleftrightarrow }&\ \ \hbox{\it 4 Dirac yinos}\  \ &\stackrel{N=2}{\longleftrightarrow } 
&\ \ \hbox{\it 5 charged spin-0 bosons}\ && \hbox{\it (type III)}
\ea \ \right\} \ \ \ \hbox{with} \ \ Z = \pm\,m_X\,,
\ee
both with the same value $\pm \,m_X$ of the central charge $Z$. The mass relation 
\be
\label{wxy}
\,m_Y^2=  \,m_W^2+m_X^2\,,
\ee
is interpreted by viewing  $m_W$ and $m_Z$ as mass parameters already present in the 5-or-6 dimensional spacetime 
as a result of the electroweak breaking.
$m_X$ is associated with extra components of the momenta along compact dimensions, $m_W^2$ and $m_X^2$ both contributing to 
the $Y^{\pm 1/3}$ mass$^2$ in 4 dimensions, according to  (\ref{wxy}) \cite{56}.

 \vspace{2mm}
This is essential when discussing $N=2$ extended supersymmetric electroweak 
or grand-unified theories, which now require {\it four rather than two spin-0 BEH electroweak doublets}, or grand-unification quintuplets
\cite{56bis,56}.

\subsection{\boldmath Electroweak breaking \,with an unbroken $\,SU(4)$ {\em electrostrong\,} symmetry, \,in 6 dimensions}
\label{subsec:extra0}

$N=2$ theories \cite{hyper,24}, and extended $N=2$ supersymmetric GUTs may be formulated from a higher (5 or 6) dimensional spacetime 
\cite{56bis,56}, with

\vspace{-5mm}
\be
V^{\hat\mu}=\ \left(\ba{c}V^\mu\vspace{1mm}\\ V^5=a\vspace{1mm}\\V^6=b \ea \right)\,.
\ee
In this higher-dimensional space, the $SU(5)$ symmetry is broken through the BEH-quintuplet v.e.v.'s,
providing in 6 dimensions equal masses to the $ Y^{\pm 1/3}$ and $W^{\mp}$ gauge fields, according to
\be
SU(5)\ \ \ \stackrel{\hbox{\footnotesize \em EW breaking}}{\longrightarrow} \ \ \ SU(4) \ \ \hbox {\it electrostrong gauge group} \ \ \hbox{in 6d spacetime,}
\ee
with
\be
SU(5)\ \ \underline{\bf 24}\ 
\left\{\ 
\ba{ccccc}
\hbox{gluons, photon,} \ X^{\pm 4/3}\ && \ \underline {\bf 15} \ &\ \hbox{\it adjoint}\ SU(4) \ \hbox{\it gauge bosons,}\!\!&\hbox{massless in 6d}\,,
\vspace{2mm}\\
Y^{\pm 1/3},\, \ W^{\mp} \ &\rightarrow\ & \  (\,\underline {\bar {\bf 4}} \, +\,\underline  {\bf 4}\,)\  &SU(4) \ \hbox{\it quartets,} \!\!
&m_Y = m_W \ \hbox{in 6d}\,,
\vspace{2mm}\\
Z\ &&\ \underline {\bf 1}\  &SU(4) \ \hbox{\it singlet,}\!\! &\hbox{$m_Z=m_W/\cos\theta\,$  in 6d}\,,
\ea \right.
\ee

\vspace{1mm}
\noindent
where $\sin^2\theta= 3/8$ at this stage.

\subsection{Grand-unification and supersymmetry breaking from extra dimensions}
\label{subsec:extra}

The extra compact space dimensions may then play an essential role in the breaking of the supersymmetry and grand-unification symmetries, 
with boundary conditions 
involving continuous or, more interestingly, discrete symmetries, including $R$-parity  and various parity-like or similar symmetries in compact space.
The breaking of supersymmetry (i.e. of the $N=2$ supersymmetry generators after reduction to 4 dimensions) 
may be obtained by identifying the action of performing a complete loop in compact space 
(e.g. a translation $x^6 \to x^6+L_6$ 
on a flat torus) with a discrete $R$-parity transformation, $R_p=(-1)^{3(B-L)}\ (-1)^{2S}$.

\vspace{2mm}
In a similar way, the breaking of the $SU(4)$ electrostrong symmetry
present in 6d 
may be obtained by identifying performing a complete loop (e.g. a translation $x^5 \to x^5+L_5$ on a torus) 
with a discrete $Z_2$ GUT-parity transformation $G_p$. This one may be defined as 
\be
\hbox{\em GUT-parity}\ \ G_p\,=\ G'\,\times\ e^{\,\frac{3}{5}\,Y}\,=\,\pm\, 1\,,
\ee
where $G'$ is a global symmetry operator commuting with $SU(5)$ and  supersymmetry. It acts on matter multiplets,
associated with quarks and leptons including mirror partners,
and BE-Higgs multiplets, in quintuplet representations of $SU(5)$.
The $X^{\pm 4/3}$ and $Y^{\pm 1/3}$ fields, and associated supersymmetry multiplets, have $G_p=-1$ and get excluded from the low-energy spectrum as a result of the compactification.
The gluon, photon, $W^\pm$ and $Z$ 
have $G_p=+1$, thus surviving as massless or light fields 
as compared to the compactification scales.
Within quintuplet BEH representations electroweak doublets have $G_p=+1$ and color triplets $G_p=-1$.
Generating $m_X$ in this way, in connection with GUT-parity, automatically solves the so-called ``doublet-triplet splitting problem''.
The $R$-parity and GUT-parity operators
$R_p$ and $G_p$ have $\pm 1$ eigenvalues, and commute.

\vspace{2mm}

A breaking of chirality in 4 dimensions is required if we are to avoid mirror quarks and leptons in the low-energy theory.
It may be obtained by considering a discrete reflexion symmetry, or rotation of $\pi$,
in the 2d compact space,  transforming $x^5,x^6$ 
into $-x^5,-x^6$ in view of identifying opposite points. 
As the $V^5=a$ and $V^6=b$ components of the $V^{\hat \mu}$ gauge fields transform with a $-$ sign, 
these extra components describing 4d fields 
associated with spin-0 gluons and photons (or more generally spin-0 electroweak gauge fields), 
with mirror-parity $M_p=-1$ and coupling ordinary particles to their mirror partners (also with $M_p=-1$),
disappear from the low-energy spectrum, as well as the mirror quarks and leptons to which they would couple.
This operation truncates away half of the states, leaving only a $N=1$ supersymmetry in the low-energy sector after reduction to 4 dimensions, without mirror particles nor spin-0 gluons 
or photons. This one is further reduced to $N=0$, i.e. no surviving supersymmetry below the compactification scale, using boundary conditions involving $R$-parity.

\vspace{2mm}
This leads to the possibility of fixing the scales associated with these breakings in terms 
of the compactification scales for the extra dimensions \cite{56}.
 With relations like, in the simplest cases of two flat extra dimensions,
\be
\label{LL}
\left\{\ 
\ba{ccccc}
m_{3/2} \!&=&\! \dis \frac{\pi}{L_6}=\, \frac{1}{2R_6}  && \hbox{(from identification \ {\em $R$-parity} $\equiv$
translation of $L_6$\,)}\,,
\vspace{2mm}\\
 m_X \!&=&\! \dis  \frac{\pi}{L_5}= \,\frac{1}{2R_5}
&&\hbox{(from identification\  {\it $GUT$-parity} $\equiv$
translation of $L_5$\,)}\, .
\ea \right.
\ee
This use of {\it discrete boundary conditions}, involving for supersymmetry $R$-parity rather than a continuous symmetry, 
allows to link rigidly these $m_{3/2}$ and $m_X$ parameters to the compactification scales. 
This is in contrast with the initial approach
of \cite{ss}, intended to generate 4d supersymmetry-breaking parameters of moderate size in an extended supergravity theory,
ignoring  all states at the compactification scale.

\subsection{Proton stability, and quark and lepton masses}

GUT-parity also leads to {\it stabilize the proton} through the mechanism of replication of quark and lepton families introduced 
for $N=2$ supersymmetry in 4 dimensions \cite{rep},
a duplication getting sufficient in 5-or-6d to constitute {\it pairs of $SU(5)$ representations with opposite $G_p$ parities}.
\vspace{-.4mm}
For example the usual $u_L$ and $\bar u_L$ fields, normally related by $SU(5)$ under which they form a $SU(4)$ 
sextet, and coupled through the $X^{\pm 4/3}$ gauge boson, are now forbidden to do so by the boundary conditions involving GUT-parity.
They can no longer be described by $SU(5)$ components of a single representation of higher-d fields,
as both need to have GUT parity $G_p=+1$ to survive at low-energy. This requires a doubling of $SU(5)$ matter representations,
not needed for the 4 quintuplet BE-Higgs representations in 6d, whose triplet components are unwanted at low energy.
Thus
\be
\hbox{\em GUT-parity} \ \ \to\ \ \hbox{\em doubling of matter representations} \ \ \to\ \ \hbox{\em stability of the proton}\,,
\ee
at least in the simplest situations considered.

\vspace{2mm}
Constructing a mass term for the matter fermions in 6 dimensions, however,  is normally impossible here, 
these being represented by {\it 6d 8-component Weyl spinors of the same chirality $-$}, as for higgsinos, owing  to supersymmetry  \cite{56}. 
Indeed gauginos, of chirality $+$, must have  6d Yukawa couplings to both higgsinos and matter fermions. 
For example the electron and mirror-electron fields are described by the 6d Weyl spinors of chirality $-$,
\be
\label{eem}
\left(\ba{c}
e_{M\,R}\vspace{-.5mm}\\  -- \vspace{-1.5mm}\\  e_L
\ea\right)\ \ \hbox{(in an electroweak doublet) \ \ and}
\ \ 
\left(\ba{c}
e_{R}\vspace{-.5mm}\\ -- \vspace{-1.5mm}\\  e_{M\,L\,} 
\ea\right) \ \  \hbox{(singlet)}\,,
\ee
verifying in particular 
\be
\label{ref}
e(x^\mu,-x^5, -x^6)\,= \, e(x^\mu,x^5, x^6)\,,\ \ \ 
e_M(x^\mu,-x^5, -x^6)\,= \,-\,e_M(x^\mu,x^5, x^6)\,,\\ \ \  \hbox{etc.}\,,
\ee
for their $M_p=+\,1$ and $-\,1$ field components, respectively.
Quark and lepton fields (and wave functions) are developed proportionally to $\cos (2\pi n_5 x_5/L_5)\,\cos (2\pi n_6 x_6/L_6)$, with $n_5$ and $n_6$ 
 integers $\geq 0$, or $\,\sin (2\pi n_5 x_5/L_5)$ $\sin(2\pi n_6 x_6/L_6)$. 
 Mirror quark and lepton fields  involve $\,\cos (2\pi n_5 x_5/L_5)$ $\sin (2\pi n_6 x_6/L_6)$ or
  $\,\sin (2\pi n_5 x_5/L_5)$ $\cos (2\pi n_6 x_6/L_6)$, as for spin-0 gluons and electroweak spin-0 bosons, etc., such states being absent  in the low-energy spectrum,
 for which  $n_5=n_6=0$.
\vspace{2mm}

To bypass this obstruction for generating quark and lepton masses
we need to connect upper to lower components, for different 6d Weyl spinors of chirality $-$\,, with right-handed upper components and left-handed lower ones 
as in (\ref{eem}). These Weyl spinors have different electroweak properties and should connect through a doublet spin-0 6d field, in an $SU(5)$ quintuplet.

\vspace{2mm}
This may be done by coupling
\vspace{-.5mm}
$\,\overline {\psi'_-}(x^{\hat\mu}) \,\Gamma_5\,\psi_-(x^{\hat\mu})$ and $\overline {\psi'_-}(x^{\hat\mu}) \,\Gamma_6\,\psi_-(x^{\hat\mu})$ 
to spin-0 quintuplets,
with their real and imaginary parts coupled to 
$\overline {\psi'_-}\,\Gamma_5\,\psi_-$ and $\overline {\psi'_-} \,\Gamma_6\,\psi_-$
very much as for the 5th and 6th components of a $V^{\hat \mu}$ gauge field.
This requires, however, abandoning Lorentz symmetry between ordinary and compact dimensions in 6d, remembering also that it
is in any case broken by the various boundary conditions used in the compactification.
The 6d Weyl matter  fields in the $\underline {\bar{\bf 5}}$ and $\underline{\bf 10}$ representations of $SU(5)$ get coupled to the four  
spin-0 fields in quintuplet representations $h_1$ and $h_2$
\vspace{-.3mm}
 (with mirror parity $M_p=+\,1$), and $h'_1$ and $h'_2$ (with $M_p=-\,1$). 
Only the former survive  at low-energy, acquiring v.e.v.'s $v_1/\sqrt 2$ and $v_2/\sqrt 2$ breaking $SU(5)$ to $SU(4)$ in 6d.
 This generates 6d mass terms such as
\be
-\,m_e\,(\,\bar e(x^{\hat\mu})  \,e(x^{\hat\mu}) + \bar e_M(x^{\hat\mu})  \,e_M(x^{\hat\mu}) \,), \,...\,,
\ee
 for charged leptons and quarks together with  their mirror and spin-0 partners, with
 \be
m_e= h_ev_1/\sqrt 2\,,\ \ m_d= h_dv_1/\sqrt 2\,,\ \ m_u= h_uv_2/\sqrt 2\,.
\ee

This also provides a way to escape the sometimes unwanted GUT mass relations $m_d=m_e,\ m_\mu=m_s,\ m_b=m_\tau$ valid at the grand-unification scale. 
 Indeed owing to the replication of quark and leptons families  \cite{rep} associated with GUT-parity,
$d$ quarks and charged leptons, with GUT-parity $+1$, now sit in different representations of $SU(5)$, together with their replicas having GUT-parity $-1$. 
$\,(d,e'^+),\,(s,\mu'^+),\,(b,\tau'^+)$ and 
$(d',e^+),\,(s',\mu^+)$, $(b',\tau^+)$ are thus associated within six quartets of the $SU(4)$ electrostrong gauge group,
only $\,e,\mu,\tau$ and $\,d,s,b$ surviving at low-energy owing to the boundary conditions involving GUT-parity.

\vspace{2mm}

$N=2$ supersymmetry gets reduced to $N=0$ in the 4d low-energy theory by the compactification process.
Heavy states carrying compact momenta remain organized  in a $N=2\,$ spectrum,  before it gets broken in the compactification process using $R$-parity.
This leads to the 4d spectrum 
\be
\left\{
\ba{ccll}
m^2\hbox{($q,\,l$, and mirrors)} \!&=\!&  m_{lq}^2 + \left(\frac{2\pi n_5}{L_5}\right)^2+ \left(\frac{2\pi n_6}{L_6}\right)^2,&\ \ n_5,n_6\ \hbox{integer}\,,
 \vspace{2mm}\\
m^2\hbox{($\tilde q,\,\tilde l$, and smirrors)}  \!&=\!&  m_{lq}^2 + \left(\frac{2\pi n_5}{L_5}\right)^2+ \left(\frac{2\pi n_6}{L_6}\right)^2,
&\ \ n_5\ \hbox{integer},\ \ n_6\ \hbox{half-integer}\,,
 \ea\right.
 \ee
 using (\ref{ref}) for counting states at each level, with e.g. two electron and two mirror electron states, etc., for each set of integers ($n_5\neq 0, \,n_6\neq 0$).

 \vspace{2mm}
 
For $n_5=0$ i.e.~below the GUT scale, the spectrum starts e.g.~with the electron at $m_e^2$, 
two selectrons and two mirror selectrons at $m^2=m_e^2+(\pi/L_6)^2$, 
 an electron and a mirror electron at $m^2=m_e^2+(2\pi/L_6)^2$ and so on, etc..
 And similarly  for all particles from the 6d spectrum already obtained for the $X,Y,W,Z,\gamma$ and gluon multiplets of supersymmetry,  combined with the compactification using $R$-parity, GUT-parity and reflexion symmetry in compact space.
(But we might also consider, to better respect the underlying extended supersymmetry,
 generating 6d masses only for quarks and leptons with $n_5=n_6=0$,
leaving (s)quarks, (s)leptons and (s)mirror particles carrying compact momenta massless in 6d.)
 
\vspace{2mm}
Two of the four 6d spin-0 doublets,
$h_1$ and $h_2$, with $M_p=+1$, remain in the low-energy spectrum, breaking spontaneously the electroweak symmetry and 
generating quark and lepton masses.
$h'_1$ and $h'_2$, with $M_p=-1$, couple quarks and leptons to their mirror partners but do not survive in the low-energy spectrum.
The 4d  theory has the same content as the Standard Model at low-energy,
but with the two spin-0 doublets $h_1$ and $h_2$ making possible the gauge/BE-Higgs unification
that is one of the most interesting features of supersymmetric theories.

\subsection{Consequences for the grand-unification and supersymmetry-breaking scales}

Independently of specific aspects on quark and lepton mass generation,  that may still be further discussed or questioned,
this indicates that supersymmetry may only show up manifestly through the presence of $R$-odd superpartners at the compactification scale, i.e.

\vspace{-5mm}
\be
m(R\hbox{\em -odd superpartners})\  \approx\ \hbox{\em compactification scale}.
\ee
This one is not necessarily directly tied to the electroweak scale, especially as {\it the electroweak breaking can be directly formulated 
in the higher 5-or-6 dimensional space-time}, independently of the compactification scale.
It may even be quite high, especially if we consider that two compactification scales of comparable order may determine 
both the supersymmetry and grand-unification scales, as hinted to by  (\ref{LL}).
This would imply 
\be
\label{susygut0}
m(R\hbox{\em -odd superpartners})\  \approx\ \hbox{\em GUT scale},
\ee
seriously decreasing the hope of finding directly superpartners very soon. On the positive side however, it would alleviate or solve the difficulties associated with flavor-changing neutral current processes 
induced by squark or slepton exchanges. And the possible stability of the proton is this framework 
might allow for the grand-unification scale to be lower than usually expected ...

\section{Back to the Supersymmetric Standard Model}
\label{sec:ssm}

\subsection{\boldmath The need for superpartners}

Let us now return to the more familiar story of simple ($N=1$) supersymmetric theories, in 4 spacetime dimensions.
Irrespectively of all the difficulties, one first had to find out which 
bosons and fermions could be related.  One may try as a warm-up exercise the tentative associations

\vspace{-5mm}
\be
\left\{\ \ \ 
\ba{ccc}
 photon\ \ &\stackrel{\hbox{ ?}}{\longleftrightarrow}&\ \ neutrino\,, 
\vspace{-.5mm}\\ 
 W^\pm\ \ &\stackrel{\hbox{ ?}}{\longleftrightarrow}&\ \  e^\pm\,,
\vspace{-.5mm}\\ 
gluons\ \ &\stackrel{\hbox{ ?}}{\longleftrightarrow}&\ \ quarks\,, \vspace{-.5mm}\\ 
& ...& \ \ \ \ \ \ \ \ 
\ea \right.
\ee

\vspace{-2mm}
\noindent
But we have no chance to realize in this way systematic associations of known fundamental bosons and fermions, 
especially as we know more fundamental fermionic field degrees of freedom, describing quarks and leptons (90), 
than bosonic ones (28 including the newly-found boson). In addition these fields have different gauge and internal 
($B$ and $L$) quantum numbers.

\pagebreak

Still the exercise of trying to relate, within a first electroweak model, 
the photon with a ``neutrino'' and the $\,W^-$ with an ``electron'', accompanyied by a
``heavy electron'' and charged BE-Higgs boson $H^-$, turned out to be very fruitful. While illustrating how far one could go when trying to relate known particles 
and the limitations of this approach, it provided through a reinterpretation of its fermions as  charginos and neutralinos
the electroweak sector of supersymmetric extensions of the standard model \cite{R}.  The initial need for a conserved quantum number carried 
by the ``lepton'' candidates led to a $U(1)_R$ symmetry acting chirally on the supersymmetry generator according to
\be
\label{Rbis}
Q\ \stackrel{R}{\rightarrow}\ e^{-\gamma_5\alpha}\ Q \ .
\ee
Both doublets $h_1$ and $h_2$ used for the electroweak breaking having $R=0$, 
this continuous $R$-symmetry survives this breaking, leading to an additive quantum number $R$ differing by $\pm 1$ unit
between bosons and fermions within multiplets of supersymmetry, gauge and BE-Higgs bosons having $R=0$, and their superpartners, 
known as gauginos and higgsinos, $R=\pm 1$.
The would-be ``neutrino'', however, uncoupled to the $Z$, 
cannot be interpreted as a $\nu_e$ or $\nu_\mu$, the $\nu_\tau$ being still unknown. It has to be viewed as a neutrino of a new type, 
a ``photonic neutrino'' that became the photino \cite{ssm}. 
The ``lepton'' candidates are thus interpreted as  {\it charginos} and {\it neutralinos}, 
providing the electroweak gauge-and-Higgs sector of the supersymmetric standard model.

\vspace{2mm}

We also have to deal with the systematic appearance 
of {\it self-conjugate Majorana fermions}, while Nature seems to know only Dirac fermions.
How can we obtain the usual Dirac fermions,
and attribute them conserved quantum numbers like $B$ and $L$\,?
This problem gets more acute as baryon number $\,B\,$ and lepton number $\,L\,$ are
carried by fundamental fermions only, quarks and leptons, 
not by bosons. This gets impossible to realize in a supersymmetric theory where bosons and fermions are related.
It also seemed to make 
supersymmetry irrelevant to the real world.
The solution consists first in accepting the existence of Majorana fermions as 
belonging to a new class of particles.
The photon 
gets associated with its own new ``neutrino'',
a {\it photonic neutrino} called in 1977 the {\it photino}; 
and  similarly for the gluons associated with {\it gluinos}, etc.\,\cite{ssm,grav}. The Majorana fermions of supersymmetry are thus  identified as gluinos
and neutralinos, or combine into charginos.
At the same time one introduces new bosons carrying baryon and lepton numbers,
known as {\it squarks and sleptons} \,--\, although this does not necessarily guarantee yet that 
$B$ and $L$ will always remain conserved at least to a sufficiently good approximation.

\vspace{2mm}

Supersymmetry thus does not relate directly known bosons and fermions.
All particles get associated with new superpartners, according to 
\vspace{-1mm}
\be
\label{relsusy}
\left\{\ 
\begin{tabular}{ccc}
{\it  known bosons } & $\longleftrightarrow$& {\it \ \ new fermions},\ \ \\[1mm]
{\it \ \ known fermions} \ \ & $\longleftrightarrow$& {\it  new bosons}.
\end{tabular} \right.
\ee
This was long mocked as a sign of  the irrelevance of supersymmetry.
But times have changed, and the same feature now gets frequently viewed, rather naively,  as an ``obvious''
consequence of the supersymmetry algebra.

\subsection{The basic ingredients}

However, even after accepting the introduction of superpartners one
still has to face another potential problem. 
With so many spin-0 particles, including squarks and sleptons
carrying $B$ and $L$, how can we get interactions 
mediated by spin-1 gauge bosons,
avoiding unwanted squark and slepton exchanges that would lead to $B$ and $L$
violations\,?
$R$-parity plays here an essential role by {\it automatically forbidding direct exchanges of squarks and leptons}
between quarks and leptons, that might otherwise lead to proton decay at a much too high rate.

\vspace{2mm}

The required ingredients for a supersymmetric extension of the Standard Model are~\cite{ssm}
\be
\left\{\ \begin{tabular}{l} 
 \ \ 1) \ $SU(3)\times SU(2)\times U(1)\,$ \it gauge superfields,
\\ [1mm]
 \ \ 2) \ \it chiral quark and lepton superfields, \  \\ [1mm] 
 \ \ 3) \ \it the two doublet BE-Higgs superfields  $H_1$  and  
$H_2$,
\\ [1mm]
 \ \ 4) \ \it a trilinear superpotential ${\cal W}_{l,q}$ responsible for $q$ and $l$ masses.\ 
\end{tabular} \right.
\ee
We must use {\it two}  doublet Higgs superfields.  
\vspace{-.8mm}
With a single one, say $H_1$, we could only construct, from 
$\,\widetilde W_{L+R}^{\,-}$ and $\,\tilde h_{1L}^{\,-}$, \,a single massive Dirac chargino  $\,\tilde h_{1L}^{\,-}+\widetilde W_{R}^{\,-}$ acquiring its mass from $<h_1^\circ>$,
\vspace{-.5mm}
\,getting stuck with a surplus massless chiral chargino $\,\widetilde W_{L}^{\,-}$.
Two doublet Higgs superfields now called 
\be
H_1=\ \left(\ba{c}H_1^0 \vspace{.5mm} \\ H_1^- \ea\right)  \ \ 
\hbox{and}\ \ \  H_2=\ \left(\ba{c}H_2^+ \vspace{.5mm} \\ H_2^0 \ea\right) \ 
\ee
are required to get {\it two} massive Dirac charginos (or actually winos) from gaugino 
($\lambda^-\!\!=\widetilde W_{L+R}^{\,-}$) 
\vspace{-.5mm}
and higgsino 
 \,($\,\tilde h_{1\,L}^{\,-}$ and $\,(\tilde h_{2\,L}^{\,+})^c$)
components.

\vspace{2mm}

To account for $B$ and $L$ conservation the superpotential $\,{\cal W}\,$
should be taken as {\it an even function of quark and lepton superfields\,} i.e. invariant under $R$-parity~\cite{ssm}.
It includes the trilinear terms
\be
\label{wlq}
{\cal W}_{l,q}\,=\, h_e \, H_1 \,.\,\bar E \,L \, +\, 
h_d \, H_1\,. \,\bar D \,Q \,-\, 
h_u \, H_2 \,.\,\bar U \,Q\,.
\ee
$H_1$ and $H_2$ are separately responsible for charged-lepton and down-quark masses, and up-quark masses, respectively, with
\be
m_e=h_e \,v_1/\sqrt 2\,,\ 
m_d=h_d\, v_1/\sqrt 2\,,\ 
m_u=h_u\,v_2/\sqrt 2\,,
\ee
and $\,\tan\beta=v_2/v_1$. This tends to favor a smaller $v_1$ as compared to $v_2$
i.e. a large $\tan\beta$, in view of the large mass of the $t$ quark as compared to $b$.

\vspace{2mm}

The superpotential ${\cal W}_{l,q}$ is also ``invariant'' under the continuous $R$ symmetry (\ref{defr0}) under which \cite{R,ssm}
\be
\label{defr2}
\left\{
\ba{ccc} 
\ \ \ V(x,\theta,\bar\theta)\, &\stackrel{R}{\rightarrow} &\, V(x,\theta\,e^{-\,i\,\alpha},\bar\theta\,e^{i\,\alpha})\,,\ \ \ 
\vspace{1mm}\\
\ \ \ H_{1,2}\,(x,\theta)\, &\stackrel{R}{\rightarrow} & \,H_{1,2}\,(x,\theta\,e^{-\,i\,\alpha})\,, \ \ \ \ \ 
\vspace{1.5mm}\\
(L,Q,\bar E,\bar D,\bar U)(x,\theta) \,&\stackrel{R}{\rightarrow} & \,e^{i\,\alpha}\ (L,Q,\bar E,\bar D,\bar U)(x,\theta\,e^{-\,i\,\alpha})\,,
\ea
\right.
\ee
so that it transforms  according to
${\cal W}_{l,q}\,(x,\theta)\,\to\,e^{2\,i\,\alpha}\ {\cal W}_{l,q}\,(x,\theta\,e^{-\,i\,\alpha})$ as in (\ref{wr0}).
The $\mu H_1H_2$ superpotential  mass term, however, does not, the higgsino mass term transforming chirally under $R$
(as a gaugino mass term but in the opposite way).
The superpotential ${\cal W}_{l,q}$ is also invariant under the extra $U(1)_A$ symmetry under which \cite{R,ssm}
\be
\label{defua}
\left\{
\ba{ccc} 
\ \ \ V(x,\theta,\bar\theta)\, &\stackrel{U(1)_A}{\rightarrow} &\, V(x,\theta,\bar\theta)\,,\ \ \ 
\vspace{1mm}\\
\ \ \ H_{1,2}\,(x,\theta)\, &\stackrel{U(1)_A}{\rightarrow} & \,e^{i\,\alpha}\ \ H_{1,2}\,(x,\theta)\,, \ \ \ \ \ 
\vspace{1.5mm}\\
(L,Q,\bar E,\bar D,\bar U)(x,\theta) \,&\stackrel{U(1)_A}{\rightarrow} & \,e^{-i\,\alpha/2}\ (L,Q,\bar E,\bar D,\bar U)(x,\theta)\,,
\ea
\right.
\ee
as in (\ref{u1a}).
\,The $\mu H_1H_2$ mass term, however, is not, the higgsino mass term transforming chirally under $U(1)_A$.

\begin{table}[tb]
\caption{Minimal particle content of the Supersymmetric Standard Model.
The gaugino fields  $\,\tilde W_3\,$ and $\,\tilde W'$  mix with the higginos
 $\,\tilde h_1^{\circ},\, \tilde h_2^{\circ}\,$ into four Majorana neutralinos; or more, as in the presence of
an extra N/nMSSM singlet $S$ coupled through a $\lambda \,H_1 H_2 \,S$ superpotential.
\label{tab:mssm}}
\vspace{0.05cm}
\begin{center}
\begin{tabular}{|c|c|c|} \hline 
&&\\ [-3.5mm]
Spin 1       &Spin 1/2     &Spin 0 \\ [.1 true cm]\hline\hline 
&&\\ [-0.3true cm]
gluons ~$g$        	 &gluinos ~$\tilde{g}$        &\\[.5mm]
photon ~$\gamma$          &photino ~$\tilde{\gamma}$   &\\ [-.5mm]
------------------&$- - - - - - - - -$&-------------------------- \\ [-4.5mm]
 

$\begin{array}{c}
\\ W^\pm\\ [1mm]Z \ \  \\ [1mm]
\\ \\
\end{array}$

&$\begin{array}{c}
\hbox {winos } \ \widetilde W_{1,2}^{\,\pm} \\ [1mm]
\,\hbox {zinos } \ \ \widetilde Z_{1,2} \\ [1mm]
\hbox {higgsino } \ \tilde h^\circ 
\end{array}$

&$\left. \begin{array}{c}
H^\pm\\ [1mm]
 H\ \\
[1mm]
h, \,A
\end{array}\ \right\} 
\begin{array}{c} \hbox {BE-Higgs}\\ \hbox {bosons} \end{array}$  \\ &&\\ 
[-.7true cm]
\hline &&

\\ [-4mm]
&leptons ~$l$       &sleptons  ~$\tilde l$ \\[0mm]
&quarks ~$q$       &squarks   ~$\tilde q$\\ [-3.5mm]&&
\\ \hline
\end{tabular}
\ec
\vspace{-2mm}
\end{table}

\noindent

\vspace{2mm}
A continuous $R$-invariance, however, would force  Majorana gluinos  to remain massless. It  must 
get reduced to $R$-parity, so that gluinos, in particular, can acquire a mass.
This also applies to the gravitino, and to gravity-induced supersymmetry-breaking terms including gaugino mass terms
\cite{grbr,cfg,grbr2}.
 (We leave aside the possibility of Dirac gluinos \cite{glu} considered in Sec.\,\ref{subsec:meta} and in theories with extra dimensions
 \cite{56} as discussed in Sec.\,\ref{subsec:extra}, with Dirac gluinos at the compactification scale, 
 making somewhat obsolete the other usual mechanisms for generating supersymmetry-breaking terms.)

\vspace{2mm}

Indeed massless or light gluinos would combine with quarks, antiquarks and gluons to form light
{\it $R$-hadrons, $R$-mesons} $\tilde g q \bar q$ and {\it $R$-baryons} $\tilde g q \bar q$ \cite{ff,glu,rh}. 
They are expected to decay into ordinary hadrons plus unobserved  neutralinos (taken at the time as photinos or very light gravitinos) 
carrying away missing energy-momentum. They were not observed, leading to consider massive gluinos, 
with the $U(1)_R$ symmetry group reduced to its discrete $Z_2$ $R$-parity subgroup, with
$R_p=(-1)^R$.
The existence of gluinos lighter than $\sim  1$ TeV$/c^2$ has now been explored at LHC, and often
excluded in this mass range in many situations of interest.

\vspace{2mm}
$R$-parity  is thus simply
\be
R_p\,=\,(-1)^R = \left\{ \ba{cc}
+1 & \hbox{\em for gauge bosons, spin-0 BEH bosons; quarks and leptons,}
\vspace{1mm}\\
-1 &\ \hbox{\em for superpartners: \ gauginos and higgsinos; squarks and sleptons.}
\ea\right.
\ee
As $\,(-1)^{2S}\equiv(-1)^{3B+L}$ for ordinary particles, it may be reexpressed as~\cite{ff}
\be
R_p\,=\,(-1)^R\, \equiv\, (-1)^{2S}\ (-1)^{3B+L}\,,
\ee
or equivalently $(-1)^{2S}\ (-1)^{3(B-L)}$. This symmetry, if conserved, requires that supersymmetric particles be produced in pairs,
their decays producing again particles of $R_p=-1$ (or an odd number of them), the ``lightest supersymmetric particle'', or LSP,
being stable. 

\vspace{2mm}

The LSP is usually taken as a neutralino, although other superpartners may also play this role.
The pair-production of supersymmetric particles \cite{ssm},  first considered for gluinos \cite{ff,rh}, sleptons \cite{ff3,sleptons} and photinos \cite{phot},
then extended to
many  processes at much higher energies~\cite{susyat,susycms,pdg}, should ultimately lead to two unobserved neutralinos, 
the famous ``missing energy-momentum'' signature often used to search for supersymmetry. 
These neutralinos may have interactions of roughly weak-interaction strength, although there magnitude depends significantly of the mass spectrum, including squark and slepton masses.

\vspace{2mm}

Once we accept that supersymmetry does not relate directly known bosons and fermions,
we have to account for the fact that superpartners have not been observed, which requires them to be sufficiently 
heavy.
If a few GeV$/c^2$ or $\,\simge 15 $ GeV$/c^2$ could be sufficient in the late seventies or early eighties \cite{rh,ff3,sleptons}, at the time of PETRA and PEP experiments,
we know now, from LEP, Fermilab then LHC experiments \cite{susyat,susycms,pdg}, that superpartners, excepted possibly
some of the neutralinos, should be rather heavy, i.e. more than $\approx 1$ TeV or so in most cases
for strongly-interacting squarks and gluinos.
Indirect constraints, discussed in other articles in this volume, may point in the direction of even heavier squarks and sleptons,
to avoid unwanted flavor-changing neutral current effects, depending also on the structure of the supersymmetry-breaking terms considered.

\vspace{2mm}

The mass splittings between squarks and quarks would be generated in global supersymmetry from the
auxiliary 
$\,<D>\,$'s \,of neutral gauge superfields~\cite{fi,R} and $\,<\!F\!>\,$'s \,of chiral ones~\cite{F1,or} most notably 
$\,H_1^{\,0}$ and $\,H_2^{\,0}\,$, in connection with the supercurrent conservation equation\,\cite{grav},
resulting in the mass sum rule  \cite{fayet79}
\be
\label{sumrule}
\overline \Sigma \ m^2(\hbox{squarks}) \ =\ \overline \Sigma \ m^2(\hbox{quarks})\ \ \ \ 
\left\{\ \ba{l}
\ \ \hbox{\it in global supersymmetry,}\vspace{.5mm}\\
\ \ \hbox{\it up to radiative corrections,}\vspace{.5mm}\\
\hbox{\small \it with $SU(3)\times SU(2)\times U(1)$ \ \it  (or $SU(5)$) gauge group.}
\ea\right. 
\ee

\vspace{-1.5mm}

\noindent
For the first family,
$\,m^2(\tilde u_1)+m^2(\tilde u_2)+m^2(\tilde d_1)+m^2(\tilde d_2)= 2\, (m_u^2+m_d^2)$, 
so that one of the squarks
should have a very small or in fact negative mass$^2$, then leading to a charge and color-breaking vacuum state as discussed in 
Sec.~\ref{sub:spec}.
Making all squarks heavy, initially done through the $<D>$ from an extra $U(1)$
with non-vanishing axial couplings \cite{ssm},
is now usually realized by generating soft supersymmetry-breaking terms from supergravity or radiative corrections, in ``gravity-mediated'' 
or ``gauge-mediated'' models~\cite{grbr,cfg,grbr2,rev4}, characterized, in the latter case, by a very light gravitino LSP.

\vspace{2mm}

Once they acquire a sizeable-enough mass as from gravity-induced gaugino mass terms,
 these weakly-interacting neutralinos\,\cite{weak} (or WIMPs, for ``weakly-interacting massive particles''), 
 stable from $R$-parity conservation, 
can annihilate sufficiently to satisfy cosmological constraints.
They became natural candidates for the non-baryonic dark matter of the Universe,
as soon as  the need for such particles became manifest \cite{dm}.
With the neutral gaugino fields  $\,\tilde W_3\,$ and $\,\tilde W'$  mixing with the higginos
 $\,\tilde h_1^{\circ},\, \tilde h_2^{\circ}\,$ into Majorana neutralinos,
according to
\be
\label{fdm}
\{ W_3,\, W';\ h_1^{\circ},\, h_2^{\circ};\ ...\, \}\  \ \ 
\stackrel{\stackrel{\hbox{\footnotesize \em SUSY}}{}}{\longleftrightarrow}\ \ \ 
{
\underbrace{\{\tilde W_3,\, \tilde W';\, \tilde h_1^{\circ},\, \tilde h_2^{\circ};\, ...\, \}}_{\hbox{\it neutralinos}}}\ ,
\ee

\vspace{-4mm}

\noindent
the association
\be 
\label{fdm2}
\hbox{\it neutral gauge ($\gamma, Z, ...$) \,and \,BEH bosons ($h,H,A$,\, ...)\,} \ \
\longleftrightarrow \ \ \hbox{\it dark matter}\ \ \ 
\ee

\vspace{1mm}

\noindent
serves as a substitute for (\ref{unif}), now relating the mediators of electroweak forces (or mass generation) to 
{\it dark matter}, rather than normal matter.
The neutralino sector, and thus the LSP which serves as a natural dark matter candidate, 
may also involve an extra higgsino ({\it singlino})\, described by a chiral singlet superfield $S$ coupled to $H_1$ and $H_2$ 
through  a $\,\lambda\ H_1H_2\,S\,$ superpotential as in the N/nMSSM, or {\it another gaugino} associated with 
an extra $U(1)$ as in the USSM.

\vspace{2mm}

The quartic interactions of the doublets $h_1$ and $h_2$ now
appear as {\it electroweak gauge interactions}, 
with a well-specified potential~\cite{R}
\be 
\label{quartic}
V_{\,\rm quartic}\, = \  \frac{g^2+g'^2}{8}\ (h_1^\dagger \,h_1-h_2^\dagger\, h_2)^2 +\, 
\frac{g^2}{2}\  |h_1^\dagger \,h_2|^2\, ,
\ee
present in all versions of the supersymmetric standard model, from the MSSM to the N/nMSSM, USSM, etc..
{\it These quartic Higgs couplings, now fixed by the electroweak gauge couplings $g$ and $g'$},
to $\,(g^2+g'^2)/8\,$ and $\,g^2/2\,$, are responsible for the mass terms $m_W^2$ and $m_Z^2$  for the charged and neutral spin-0 BEH fields 
$w^\pm$ ($\equiv H^\pm$) and $z$ 
that appear as spin-0 partners of the $W^\pm$ and $Z$.
This is at the root of the gauge/BE-Higgs unification described in Sec.\,\ref{sec:gh}, 
and resulting mass  relations.

\subsection{\boldmath  \vspace{1mm}
Turning $\mu$ into a dynamical superfield variable in superspace}

One can introduce  as in \cite{R} ($\mu$ being then called $m$) the $ \mu\,H_1 H_2$ superpotential term,
leading to 
 the mass terms
\be 
\label{mumass}
V_{\mu}\, = \  \mu^2\ (h_1^\dagger \,h_1 +h_2^\dagger\, h_2)\, .
\ee
But even in the presence of the weak-hypercharge $\xi D'$ term splitting the $h_1$ and $h_2$ mass$^2$ terms apart from $\mu^2$ we cannot get non-vanishing 
v.e.v.'s for both $h_1$ and $h_2$ but only for one of them, the other being ``inert'', getting stuck with a massless chargino.
Adding a soft term proportional to $h_1 h_2$  that would cure the problem but at the price of making the theory non-supersymmetric was not considered, at the time, as a valid option.
The $\mu\,H_1 H_2$  term is also not invariant under the continuous $R$ symmetry, 
which would require $\mu$ to transform according to $\mu\to e^{2i\alpha} \mu$. The $\mu$ parameter was thus 
made dynamical in superspace  by being promoted to a superfield variable $\,\mu(x,\theta)$  through the 
introduction of the extra singlet $S$, with the replacement 
\be
\mu\ \to \ \mu(x,\theta) \,=\, \lambda \, S(x,\theta)\, \ \ \ \hbox{\small so that}\ \ \ \mu\,H_1 H_2\ \to \ \lambda\,H_1 H_2\,S\,.
\ee
The resulting superpotential ${\cal W}$ reads, when $R$-invariance is not required,
\be
\label{tril}
{\cal W} \ = \ \lambda\,H_1 H_2\,S \ [ + \,\mu\,H_1 H_2] \, + f(S)  + {\cal W}_{lq} \ ,
\ee
with 
\be
\label{fS}
f(S)\ =\ \frac{\kappa}{3}\ S^3\,+\,\frac{\mu_S}{2}\ S^2\,+\,\sigma\,S\,,
\ee
as in the general NMSSM. 

\vspace{2mm}

The additional requirement of $R$-invariance (i.e.~\,$U(1)_R$) was initially intended to get an additive quantum number $R$ carried by Dirac 
fermions, leading to {\it Dirac} neutralinos and charginos. 
Demanding $R$-invariance (acting as in (\ref{trhs},\ref{defr2})) reduced the singlet superpotential $f(S)$ 
 to the sole linear term $\,f(S)=\sigma\,S\,$ of the nMSSM, with
 \be
{\cal W} \ = \ \lambda\,H_1 H_2\,S +\ \sigma\,S\,  +\, {\cal W}_{lq} \,.
\ee
The extra trilinear and bilinear terms $\frac{\kappa}{3}\, S^3$ and $\frac{\mu_S}{2}\,S^2$, excluded 
by the continuous $R$ as well as the $\mu$ term, possibly regenerated from $<S>$ according to 

\vspace{-6mm}

\be
\mu_{\rm eff}\ =\ \lambda\,<S>\ ,
\ee
may be reintroduced  if we do not insist on this symmetry (as required for Majorana gluino and gravitino masses).
 This leads to the chargino (wino) mass matrix
\be
{\cal M}\  =\  \left( \ba{cc} m_2\! &\!m_W\,\sqrt 2\,\sin\beta \vspace{1mm}\\
m_W\,\sqrt 2\,\cos\beta\!&\!\mu_{\rm eff} \ea\right)\ ,
\ee
where non-diagonal terms respect the continuous $R$  while diagonal ones violate it by $\Delta R =\pm 2$.
Having both $\,m_2\,$ and $\,\mu_{\rm eff}\,$ different from zero is essential to have
both charginos
(winos)  heavier than $\,m_W$, as now necessary \cite{cfg,pdg}.

\vspace{2mm}
The $\mu$  parameter, although ``supersymmetric''
(and possibly regenerated from  $<S>$ as $\mu_{\rm eff}$)
may be protected from being too large as it comes in violation of both the continuous $R$-symmetry and chiral $U(1)_A$ 
acting according to \cite{R} 

\vspace{-6mm}
\be
H_{1,2}\,(x,\theta)\ \stackrel{R}{\rightarrow}\ H_{1,2}\,(x,\theta\,e^{-\,i\,\alpha})\,,\ \ \ 
H_{1,2}(x,\theta)\ \stackrel{U_A}{\rightarrow}\,e^{i\alpha}\,H_{1,2}(x,\theta)\,.
\ee
The same continuous $R$ symmetry may prevent gluino and other gaugino masses from being too large. It is even so efficient in doing so that it makes difficult 
to generate significant gluino masses 
from radiative corrections involving messenger quarks \cite{glu} 
unless these are taken really very heavy (introducing a new scale much larger than electroweak scale).
$\mu$ and gaugino mass parameters ($m_{1/2}$) may then naturally be of the same order,
i.e.
\be
\label{mug}
\mu\ \,(\hbox{or }\ \mu_{\rm eff})\, \approx\,m_{\rm gauginos}\,\approx\,m_{\rm gravitino}\ \ \ \hbox{in view of $R$ symmetry}\,.
\ee

\vspace{1mm}
A further advantage of the dynamical change $\,\mu\to \lambda\,S$ comes with the introduction of new quartic couplings in the potential, independent of gauge couplings. The doublet mass terms for $h_1$ and $h_2$ get replaced by $\lambda^2\,|s|^2$
$(h_1^{\,\dagger}h_1+h_2^{\,\dagger}h_2)$, \,with
the quartic potential (\ref{quartic}) replaced as in \cite{R} (where $\lambda^2$ was called $h^2/2$, 
\,with $\,\tan\beta=v_2/v_1\equiv \tan\delta=v'/v"$) \,by
\be 
\label{quartic3}
V\,= \ \frac{g^2+g'^2}{8}\ \ (h_1^\dagger \,h_1-h_2^\dagger\, h_2)^2\, +\, 
\frac{g^2}{2}\ \ |h_1^\dagger \,h_2|^2\ +\ \left|\,\lambda\, h_1 h_2+\frac{\partial f(s)}{\partial s}\,\right|^2\, +\ \lambda^2\,|s|^2\ (h_1^{\,\dagger}h_1+h_2^{\,\dagger}h_2)\ +\ ...\ \ .
\ee
The extra 
quartic coupling 
\be
\label{quartic2}
V_{\,\rm quartic}^{\,\lambda}=\,\lambda^2\ |h_1\,h_2|^2\, +\, \lambda^2\,|s|^2\ (h_1^{\,\dagger}h_1+h_2^{\,\dagger}h_2)\,,
\ee
comes in addition to the $\,(g^2+g'^2)/8\,$ and $\,g^2/2$ terms in (\ref{quartic}) and is essential to make all spin-0 BE-Higgs bosons sufficiently heavy, now leading to
\be
\label{hZ2}
m_h^{\,2}\ \,\leq\ \,
m_Z^{\,2}\, \cos^2 2\beta\, +\, \hbox{\small $\dis \frac{\lambda^2\,v^2}{2}$}\, \sin^2 2\beta\ \ \
(\,+\ \hbox{\small rad.~corr.} \,)\,,
\ee
leading to an upper bound larger than $m_Z$ (+ radiative corrections), for $\lambda>\sqrt {(g^2+g'^2)/2}$ so that $\lambda v/\sqrt 2 > m_Z$, independently of $\tan\beta$.
This makes much easier to obtain a lightest spin-0 mass of 125 GeV$/c^2$  in the N/nMSSM, without having to rely on large contributions from radiative corrections involving 
very heavy stop quarks. If sizeable enough the coupling $\lambda$ of the trilinear superpotential $\lambda\,H_1H_2 S$ also makes it easier to consider smaller $\tan\beta\,$ that may be $\approx 1$, in contrast with the MSSM which tends to require large $\tan\beta$.

\section{Conclusions and perspectives}

Our first aims were to understand how supersymmetry may allow for a description of fundamental particles and interactions, and discuss 
resulting  implications. 
At the time weak interactions through neutral currents were just recently discovered, the structure of the weak neutral current unknown,  the $W^\pm$ and $Z$ hypothetical, with a lower limit 
on $m_W$ as low as about 5 GeV$/c^2$ in 1974.  No fundamental spin-0 boson was known, and many physicists did not really believe in their existence, preferring to think of spontaneous symmetry breaking as induced by v.e.v.'s for bilinear products of fermion fields.
Lots of efforts were made developing theories trying to avoid such unwanted fundamental spin-0 fields and particles  \cite{tc,tc2}.

\vspace{2mm}
In contrast supersymmetric theories,  leading to superpartners with 
extra charged and neutral \hbox{BE-Higgs} bosons from the 
two electroweak spin-0 doublets \cite{R,ssm,ff,fayet79},
provide a natural framework for fundamental spin-0 fields  
by relating them to chiral  spin-$\frac{1}{2}$ ones (and spin-1 fields as we saw), so that
 \be
 \hbox{\it spin-0 fields acquire, within supersymmetry,  
 equal dignity as spin-$\frac{1}{2}$ ones}.
 \ee
Spin-0 mass parameters are equal, up to supersymmetry-breaking effects, to spin-$\frac{1}{2}$ ones,
free from quadratic divergences,  all benefiting from remarkable non-renormalisation properties \cite{iz}.

\vspace{2mm}
The electroweak scale is the most natural one where to search for supersymmetric particles, 
with supersymmetry usually expected to show up a not-too-high scale, now typically considered  to be $\simle$ a few TeV at most.
But no supersymmetric particle has shown up yet,  with lower limits on squarks and gluinos often reaching  $\simeq $ TeV scale \cite{susyat,susycms}.
Still {\it we keep hoping that the next round of LHC experiments will lead to a direct discovery of the new superpartners}.
In between, the finding at CERN of a new particle likely to be a  spin-0 BE-Higgs boson 
\cite{hat,hcms,higgs,higgs2} seems to indicate that we are indeed on the right track, in contrast with other anticipations 
about the nature of electroweak breaking.

\vspace{2mm}
At which energy scale we may expect to find supersymmetric particles depends on the magnitude of the parameters describing 
supersymmetry breaking, in connection with the amount of breaking of the continuous $R$ symmetry.
Gaugino mass parameters ($m_{1/2}$ i.e.~$m_1,m_2,m_3$) break both supersymmetry and $R$ symmetry,  
while the higgsino mass parameter $\mu$ (or $\mu_{\rm eff}$ if regenerated from $<S>$) also comes in violation of $R$ symmetry, 
making natural for them to be of the same order.

\vspace{2mm}

Squark and slepton supersymmetry-breaking mass$^2$ parameters ($m_\circ^2$)  may also be of this same order, as frequently considered. 
Or they could be significantly larger as they do not violate $R$ symmetry, leading to consider situations such as 

\vspace{-6.5mm}
\be
\label{mug2}
m(\tilde q,\tilde l) \ \approx \ m_\circ \ \left\{\ba{c}  \gg\ ?\vspace{.5mm}\\
\approx \ ? \ea\right\} \ 
\,m_{\rm gauginos}\,\approx\,\mu_{\rm eff}\,\simge\ \hbox{a few $m_W$ to TeV}\,.
\ee
One may also consider other situations, e.g.~with moderate $m_\circ$ as compared to very large $m_{3/2}$ and $\mu_{\rm eff}$.
 What should be the mass scale for the new particles,
if not  $\sim$ TeV scale as was commonly expected, remains an open question. 

\vspace{2mm}
As supersymmetry breaking and electroweak breaking are in general {\it two independent phenomena}, we should take seriously the possibility 
that their breaking scales be of different magnitudes.
The supersymmetry-breaking scale could then be significantly larger, {\it especially if a new  physical phenomenon not directly related to electroweak breaking is involved in this process}, such as the compactification of an extra dimension.

\vspace{2mm}
This could fix the supersymmetry-breaking scale in terms of the compactification scale ($\approx \hbar/Lc$) using $R$-parity and other discrete symmetries for the boundary conditions in compact space. Identifying 
\be
\hbox{\em performing a complete loop in compact space}\ \ \equiv\ \  R\hbox{\em -parity transformation},
\ee
we get relations  like $m_{3/2} \approx \pi/L$ (or $1/(2R)$), in the simplest case  \cite{56}.
We may then face the eventuality that superpartner masses be considerably larger than the presently accessible $\,\approx$ TeV scale, 
especially if the compactification of extra dimensions also sets the scale for grand-unification breaking.
This may tell us that supersymmetry should only show up manifestly through the presence of $R$-odd superpartners at the compactification scale, i.e.
\be
m(R\hbox{\em -odd superpartners})\  \approx\ \hbox{\em compactification scale\,?}
\ee

\vspace{.5mm}
 This one is not necessarily directly tied to the electroweak scale, especially as {\it the electroweak breaking can be 
 directly formulated in the higher 5-or-6 dimensional spacetime} (where it leaves an {\it electrostrong symmetry} unbroken), independently of the compactification scale.   It may be quite high, especially if two similar compactification scales determine 
both the supersymmetry and grand-unification scales, as hinted to by  (\ref{LL}). This would imply 
\be
\label{susygut}
m(R\hbox{\em -odd superpartners})\  \approx\ \hbox{\em GUT scale\,??}
\ee
with the further possibility that the GUT scale be lower than usually considered, in connection with the possible stability 
of the proton associated with GUT-parity.

\vspace{3mm}

Fortunately, supersymmetric theories also lead to gauge/BE-Higgs unification  by providing spin-0 bosons as extra states for spin-1 gauge bosons within massive gauge multiplets, in spite of their different gauge symmetry properties
(and independently of extra dimensions) \cite{R,gh}.
Massive gauge superfields now describe spin-0 BE-Higgs bosons, next to massive spin-1 gauge bosons.
In particular, the 125 GeV$/c^2$ boson recently observed at CERN may also be interpreted, up to a mixing angle induced by supersymmetry breaking, as
the spin-0 partner of the $Z$ under {\it two} supersymmetry transformations,
\vspace{-4mm}

\be
\label{gh3}
\hbox {\framebox [8cm]{\rule[-.35cm]{0cm}{.95cm} $ \dis
\hbox{\em spin-1} \ \,Z\ \  \ \stackrel{SUSY}{\longleftrightarrow }\ \ \stackrel{SUSY}{\longleftrightarrow }\  \ \hbox{\em spin-0 \,BEH boson}\,,
$}}
\ee
providing the first example of two fundamental particles of different spins  related by supersymmetry. {\it Supersymmetry may thus be tested in the gauge-and-BE-Higgs sector\,} at present and future colliders, in particular through the properties of the new boson, even if $R$-odd supersymmetric particles were 
still to remain out of reach for some time.

\pagebreak

 \end{document}